\documentclass{article}

\usepackage{microtype}
\usepackage{graphicx}
\usepackage{subfigure}
\usepackage{booktabs} 

\usepackage{hyperref}

\usepackage[accepted]{icml2025}

\usepackage{amsmath}
\usepackage{amssymb}
\usepackage{mathtools}
\usepackage{amsthm}
\usepackage{enumitem} 
\setlist[itemize]{topsep=2pt, partopsep=0pt, itemsep=2pt, parsep=0pt}

\usepackage[capitalize,noabbrev]{cleveref}

\theoremstyle{plain}
\newtheorem{theorem}{Theorem}[section]
\newtheorem{proposition}[theorem]{Proposition}

\theoremstyle{definition}

\theoremstyle{remark}

\usepackage[textsize=tiny]{todonotes}

\usepackage{ulem}

\usepackage{soul}

\icmltitlerunning{Low-Rank Agent-Specific Adaptation (LoRASA) for Multi-Agent Policy Learning}

\begin{document}

\twocolumn[
\icmltitle{Low-Rank Agent-Specific Adaptation (LoRASA) for Multi-Agent Policy Learning}



\icmlsetsymbol{equal}{*}

\begin{icmlauthorlist}
\icmlauthor{Beining Zhang}{equal,yyy}
\icmlauthor{Aditya Kapoor}{equal,yyy}
\icmlauthor{Mingfei Sun}{yyy}
\end{icmlauthorlist}

\icmlaffiliation{yyy}{Department of Computer Science, University of Manchester, Manchester, United Kingdom}

\icmlcorrespondingauthor{Beining Zhang}{felix.zbn@gmail.com}
\icmlcorrespondingauthor{Aditya Kapoor}{aditya.kapoor@postgrad.manchester.ac.uk}

\icmlkeywords{Machine Learning, ICML}

\vskip 0.3in
]




\printAffiliationsAndNotice{\icmlEqualContribution} 

\begin{abstract}
Multi-agent reinforcement learning (MARL) often relies on \emph{parameter sharing (PS)} to scale efficiently. However, purely shared policies can stifle each agent’s unique specialization, reducing overall performance in heterogeneous environments. We propose \textbf{Low-Rank Agent-Specific Adaptation (LoRASA)}, a novel approach that treats each agent’s policy as a specialized “task” fine-tuned from a shared backbone. Drawing inspiration from parameter-efficient transfer methods, LoRASA appends small, low-rank adaptation matrices to each layer of the shared policy, naturally inducing \emph{parameter-space sparsity} that promotes both specialization and scalability. We evaluate LoRASA on challenging benchmarks including the StarCraft Multi-Agent Challenge (SMAC) and Multi-Agent MuJoCo (MAMuJoCo), implementing it atop widely used algorithms such as MAPPO and A2PO. Across diverse tasks, LoRASA matches or outperforms existing baselines \emph{while reducing memory and computational overhead}. Ablation studies on adapter rank, placement, and timing validate the method’s flexibility and efficiency. Our results suggest LoRASA’s potential to establish a new norm for MARL policy parameterization: combining a shared foundation for coordination with low-rank agent-specific refinements for individual specialization.
\end{abstract}

\section{Introduction}
\label{sec:introduction}

A canonical paradigm in MARL is \textbf{Centralized Training and Decentralized Execution (CTDE)}~\citep{CTDE_1, CTDE_2, CTDE_3, CTDE_4, CTDE_5}, where agents learn with access to global information but execute policies independently. Within CTDE, a standard approach is \textbf{parameter sharing (PS)}~\citep{Gupta2017CooperativeMC, Chu2017PSDPG, terry2020revisiting}, which significantly cuts down on resource requirements by training a single policy network for all agents.

Despite its efficiency, PS can compromise the specialized behaviors needed in heterogeneous or role-based scenarios~\citep{christianos2021SePS, AdaPS}. Simple fixes, such as tagging states with agent identifiers~\citep{terry2020revisiting, CTDE_4, Gupta2017CooperativeMC}, rarely capture deeper skill differences ~\citep{christianos2021SePS, AdaPS}. Merely appending an ID to observations seldom suffices to uncover such diverging policies—an agent must not only “know” it has a particular identity but also adapt its policy to exploit that identity effectively. Furthermore, even in homogeneous scenarios, such as StarCraft with agents of the same unit type, we empirically found that agents also require non-identical behaviors, see Sec ~\ref{subsec:heterogeneous_nature}.

To enable better heterogeneous behaviors in multi-agent systems, researchers have explored approaches like selective parameter sharing (SePS)~\citep{christianos2021SePS} and adaptive parameter sharing (AdaPS)~\citep{AdaPS, PSNetworkPruning}. SePS clusters agents based on behavioral similarities, assigning a shared policy network to each cluster, while AdaPS dynamically selects specialized subnetworks from a shared architecture. However, SePS often struggles in dynamic environments where agent roles evolve or unique edge cases arise, as its static clustering framework cannot adapt to changes. Similarly, AdaPS can over-prune critical parts of the shared network, limiting agents’ ability to leverage common knowledge in unforeseen situations. This lack of adaptability can significantly impact performance in complex environments where flexibility is paramount. For example, in disaster response scenarios, drones performing routine tasks like surveying may be effectively managed by SePS or AdaPS, but these methods often fail to address rare, specialized tasks such as hazardous material containment or rescue operations, where more nuanced specialization is required.

\begin{figure}[!ht]
\begin{center}
    \includegraphics[width=\columnwidth]{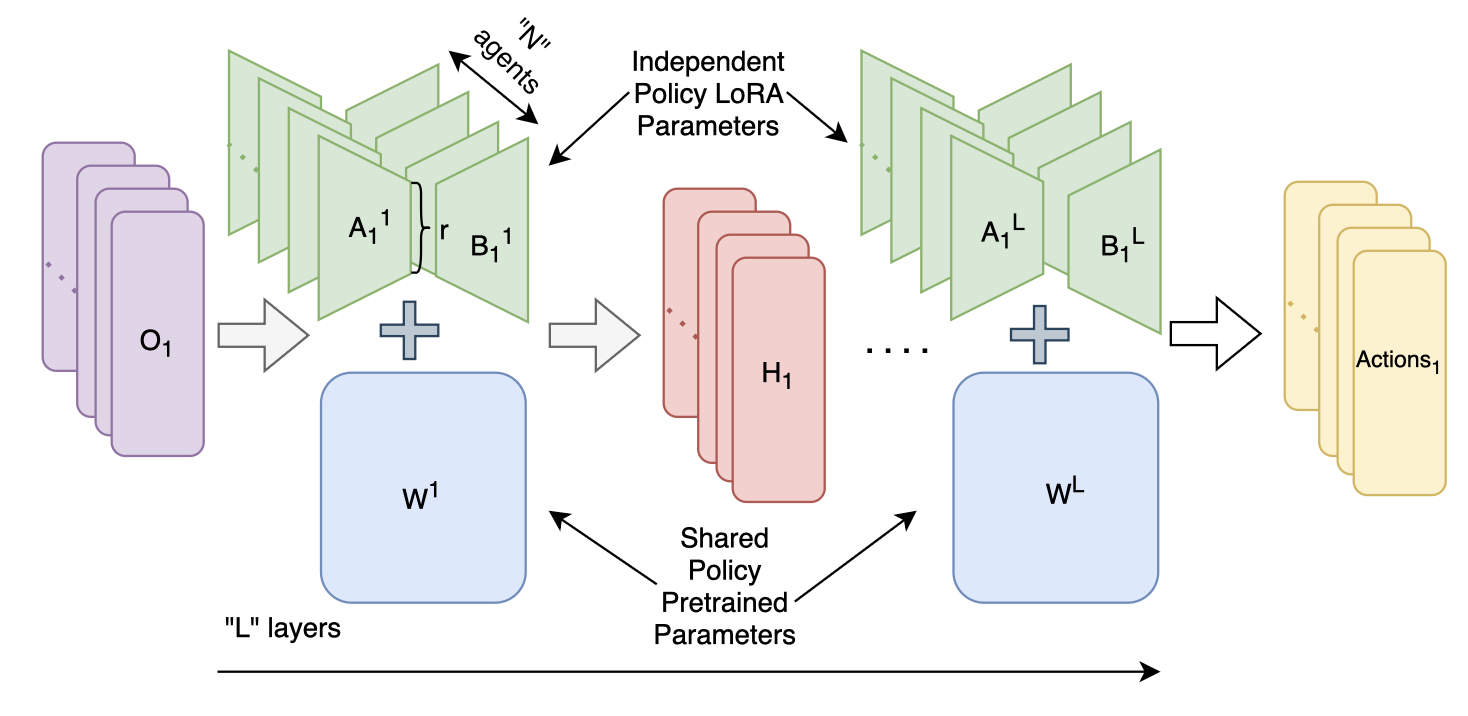}
    \caption{Overview of LoRASA framework.}
    \label{fig:LoRASA_framework}
\end{center}
\vspace{-15pt}  
\end{figure}

Fully distinct policies, such as those used in Non-Parameter Sharing (NPS), assign unique parameters to each agent~\citep{HAPPO, A2PO}. While this allows for full specialization, it forces each neural network to learn similar representations independently, often with limited data per agent, leading to sample inefficiency. Moreover, the approach is computationally and memory-intensive, making it impractical for large number of agents (see time and memory requirement of NPS in Figure~\ref{fig:resource_requirements}).

In light of these challenges, there is a strong incentive to develop methods that preserve the efficiency and coordination benefits of parameter sharing while allowing agents to specialize effectively. In this paper, we introduce a novel perspective on MARL by framing it as a multi-task learning (MTL) problem, where each agent's policy is treated as a distinct task requiring specialized behavior. Unlike previous multi-task MARL approaches that aim to generalize across similar tasks to facilitate adaptation~\citep{wang2023multitaskmultiagentsharedlayers, omidshafiei2017deepdecentralizedmultitaskmultiagent, MultiTaskMARL, zhang2024hybridtrainingenhancedmultitask}, our approach focuses on fostering diverse agent behaviors within a unified multi-agent objective. For instance, in a disaster-stricken city, rescue robots share the common skill of navigating debris-filled environments but specialize in tasks like clearing rubble, delivering medical supplies, or locating survivors. These distinct roles require specialized policies that cannot be effectively captured by a single shared network or simple agent identifier augmentations.

To address this need for efficient specialization, we propose \textbf{Low-Rank Agent-Specific Adaptation (LoRASA)}, refer Figure~\ref{fig:LoRASA_framework}, inspired by \textbf{LoRA}~\citep{hu2021lora}, a parameter-efficient fine-tuning method originally developed for large-scale natural language models. LoRA introduces lightweight, low-rank adaptation matrices that are added to pretrained weights, enabling task-specific refinements while preserving the core shared knowledge. Extending this concept to MARL, LoRASA fine-tunes the shared policy with minimal overhead by constraining adaptations to a low-rank subspace. This induces \emph{parameter-space sparsity} (Sec~\ref{paragraph:sparsity_analysis}), allowing each agent to specialize without the computational and memory burdens of assigning unique, full-rank parameters (see plots in Figure~\ref{fig:resource_requirements} for further evidence). 

Our work makes the following main contributions:
\begin{itemize}
    \item \textbf{LoRASA} (in Sec~\ref{sec:methodology}): We introduce a novel low-rank adaptation mechanism for MARL, positioning the problem as a \emph{multi-task fine-tuning} scenario to achieve both scalability and agent-specific specialization.

    \item \textbf{Comprehensive Empirical Evaluation} (refer Sec~\ref{subsec:results}): On two challenging benchmarks—StarCraft Multi-Agent Challenge (SMAC)~\citep{SMAC} and Multi-Agent MuJoCo (MAMuJoCo)~\citep{HAPPO}—LoRASA consistently matches or outperforms other parameter-sharing variants based on strong baselines (MAPPO, A2PO) while utilizing fewer resources. Our extensive ablation studies on adapter rank, fine-tuning timing, and layer-wise placement provide actionable guidelines, reinforcing LoRASA’s practicality for real-world deployment.

    \item \textbf{Parameter-Space Sparsity and Heterogeneous Behaviors} (see Sec~\ref{subsec:heterogeneous_nature}): LoRASA leverages parameter-space sparsity through low-rank updates, enabling agents to exhibit diverse, specialized behaviors while retaining the efficiency and coordination benefits of a shared policy.
\end{itemize}

Together, these contributions highlight a paradigm shift in MARL from rigid, fully shared or fully distinct policies to a flexible, low-rank adaptation framework. LoRASA enables agent-specific specialization while preserving the coordination benefits of a shared policy, offering a scalable and efficient solution for diverse and large-scale MARL applications that demand both flexibility and resource efficiency.

\section{Methodology}
\label{sec:methodology}

\subsection{Preliminaries}
\label{subsec:preliminaries}

\paragraph{Multi-Agent Reinforcement Learning (MARL).}
We consider cooperative MARL problems modeled as \textbf{Partially Observable Markov Games (POMGs)}~\citep{Kuhn, Shapely, Boutilier}, where each agent $i$ observes $o_{i,t} \in \mathcal{O}_i$, selects an action $a_{i,t} \in \mathcal{A}_i$ according to its policy $\pi_i(a_{i,t}\mid o_{i,t};\theta_i)$, and receives a \emph{shared} reward $r_t$ at time $t$. The objective is to maximize $E\big[\sum_{t=0}^\infty \gamma^t r_t\big]$, focusing on cooperative tasks.

\paragraph{Centralized Training \& Decentralized Execution (CTDE).}
In this work, we adopt \textbf{CTDE}~\citep{CTDE_1, CTDE_2, CTDE_4}, where $\mathcal{N}$ agents train with access to global information (joint observations, rewards) yet execute independently based on local observations. This setup is a natural fit for real-world multi-agent scenarios demanding high scalability and local autonomy. Under CTDE, the joint policy $\Pi$ factorizes as
\[
\Pi(a\mid o) \;=\;\prod_{i\in \mathcal{N}} \pi_i(a_i \mid o_i;\theta_i).
\]
In PS approaches, $\theta_i=\theta_\text{shared}$ for all $i\in \mathcal{N}$, while in NPS, each agent has distinct parameters. Our method, LoRASA, stands at an \emph{intermediate} point, blending the resource-efficiency of PS with the flexibility of NPS.

\paragraph{Low-Rank Adaptation (LoRA)~\citep{hu2021lora}.}
LoRA was introduced for parameter-efficient fine-tuning in large-scale language models. It adds a low-rank update $\delta W = AB^\top$ to each weight matrix $W$, where $A\in\mathbb{R}^{d\times r}$ and $B\in\mathbb{R}^{k\times r}$ with $r \ll \min(d,k)$. Critically, only $A$ and $B$ are trained, while $W$ remains fixed. In our setting, we treat the shared policy as \emph{pretrained} and each agent’s specialized adaptation as a separate \emph{task}. Thus, LoRA naturally encodes agent-specific deviations from a common baseline \emph{without} replicating entire networks.

\subsection{LoRASA: Low-Rank Adaptation for MARL}
\label{subsec:framework_overview}

\paragraph{Theoretical and Conceptual Insights.}
Recent studies in deep reinforcement learning suggest that the effective dimensionality of learned policies can be much lower than the total parameter count~\citep{remman2024discoveringbehavioralmodesdeep, Sun2019RealtimePD, ensemble_policy_distillation, schneider2024identifyingpolicygradientsubspaces}. In cooperative MARL, agents often assume distinct roles (e.g., scouting vs.\ attacking), indicating these policy variations lie in a smaller subspace of the full parameter space~\citep{wadhwania2019policydistillationvaluematching}. By restricting agent-specific updates to an \( r \)-rank matrix, LoRA formally encodes each agent’s deviations from a shared backbone within this lower-dimensional subspace. This design not only retains the bulk of the pretrained policy’s knowledge but also efficiently captures heterogeneous behaviors without duplicating entire networks.

\begin{proposition}
Assume that in a cooperative multi-agent reinforcement learning (MARL) setting, the agent-specific parameter deviations lie within or near an \( r \)-dimensional affine subspace of the full parameter space. Then, applying a rank-\( r \) low-rank adaptation (LoRA) to the shared backbone's weights can approximate the optimal agent-specific policies with a bounded error in the least-squares sense.
\end{proposition}

This proposition is supported by the Eckart-Young-Mirsky theorem~\citep{Eckart1936TheAO, Hiriart-Urruty_Le_2013}, which states that the best rank-\( r \) approximation of a matrix minimizes the Frobenius norm of the approximation error. Confining agent-specific offsets to a rank-$r$ subspace thus balances scalability and expressiveness: each agent can specialize sufficiently to capture its role-specific deviations while still sharing the bulk of learned features. In practice, this translates to improved scalability, merging the resource efficiency of parameter sharing with the fine-grained specialization of non-parameter sharing in a single low-rank framework.

\paragraph{Weight Parameterization in the Actor Network.}
Consider a recurrent actor network with fully connected (FC) layers and a recurrent unit (GRU~\citep{GRU} or LSTM~\citep{LSTM}). Let $\theta^\ell \in \mathbb{R}^{d_\ell \times k_\ell}$ be the weight matrix at layer $\ell$. We add a low-rank adaptation \(\delta \theta^\ell = A^\ell B^{\ell\top}\), where \(A^\ell\in\mathbb{R}^{d_\ell\times r}\) and \(B^\ell\in\mathbb{R}^{k_\ell\times r}\). These matrices are trained specifically for each agent, while the shared backbone \(\theta^\ell\) remains frozen. We emphasize linear transformations in the recurrent pathway (input-to-hidden and hidden-to-hidden), leaving biases and layer-norm parameters fixed for simplicity. Nonetheless, even applying LoRA solely to linear transformations gives agents ample capacity to adapt their recurrent dynamics.

\paragraph{Action Spaces and Final Layer Adaptations.}
For continuous and constrained action spaces, the actor network outputs the mean and log-std of a squashed Gaussian distribution~\citep{SAC}. We apply LoRA to the weight matrices of the final fully connected (FC) layers responsible for generating both the mean and the log-std. This allows each agent to tailor its exploration strategy through agent-specific low-rank adaptations without duplicating entire networks. For discrete action spaces, we apply LoRA to the final FC layer that produces action logits, enabling agent-specific adjustments to discrete action probabilities.

By focusing LoRA on these output layers, agents can refine their decision-making to match specialized roles (e.g., scouting vs.\ attacking) while maintaining the efficiency and coordination benefits of a shared policy backbone.

\subsection{Training Procedure}
\label{subsec:training_procedure}

LoRASA consists of two main steps: \textbf{Shared Policy Pretraining} for learning shared knowledge and \textbf{LoRA Fine-Tuning} for agent-specific specialization.

\paragraph{Phase 1: Shared Policy Pretraining.}
We first train a single shared policy \(\theta_\text{shared}\) using a standard multi-agent reinforcement learning (MARL) algorithm (e.g., MAPPO, A2PO). During this phase, the system behaves like a PS method, where all agents rely on the same policy. To evaluate the robustness of the shared policy, we track key performance metrics such as \emph{cumulative returns} and \emph{win rates}. These metrics quantify the policy's ability to exhibit effective behaviors across tasks. Once \(\theta_\text{shared}\) shows consistent improvement and meets predefined performance thresholds, we consider it sufficiently trained for downstream adaptation. At this point, we transition to fine-tuning, allowing agents to specialize their policies while retaining shared knowledge.

\paragraph{Phase 2: LoRA Fine-Tuning.}
In this phase, we introduce LoRA adapters \(\{A_i^\ell, B_i^\ell\}\) for each agent \(i\), while keeping the shared policy \(\theta_\text{shared}\) frozen:
\begin{equation}
    \forall \ell: \quad \theta_i^\ell = \theta_\text{shared}^\ell + A_i^\ell B_i^\ell, \quad A_i^\ell \in \mathbb{R}^{d \times r}, \quad B_i^\ell \in \mathbb{R}^{r \times k}. \nonumber
\end{equation}
The \textbf{low-rank dimension} \( r \) controls the level of specialization:
\begin{itemize}
    \item \textbf{Larger \( r \)} enables more expressive adaptations but increases resource costs.
    \item \textbf{Smaller \( r \)} keeps agents closer to the shared policy, ensuring efficiency.
\end{itemize}

Tuning \( r \) balances \textbf{expressivity vs.\ efficiency}—higher \( r \) approaches NPS-like fine-tuning, while lower \( r \) retains parameter-sharing benefits. During fine-tuning, agents update only \(\delta\theta_i^\ell = A_i^\ell B_i^\ell\) using their own trajectories, allowing specialization atop a shared backbone. This framework \textbf{merges PS’s scalability with NPS’s adaptability}, achieving specialization without excessive overhead.

\subsection{Algorithms for LoRASA}
\label{subsec:algos}

\paragraph{Algorithmic Details.}
Algorithms~\ref{alg:pretraining}, \ref{alg:finetuning}, and \ref{alg:inference} outline our approach. In \textbf{Algorithm~\ref{alg:pretraining}}, we train the shared policy exactly as in standard CTDE-based MARL. \textbf{Algorithm~\ref{alg:finetuning}} then enables agent-specific specialization by updating LoRA adapters for each agent’s actor network. Finally, \textbf{Algorithm~\ref{alg:inference}} merges the LoRA updates into the backbone weights at execution time for efficient inference.

\paragraph{Choice of Baseline Algorithms (MAPPO and A2PO).}
We implement LoRASA on top of two distinct CTDE actor-critic methods: MAPPO~\citep{MAPPO} extends PPO to multi-agent settings with a centralized critic and typically uses shared policy parameters, while A2PO~\citep{A2PO} sequentially updates each agent’s policy, mimicking NPS. By applying LoRASA to both methods, we illustrate how a low-rank adaptation framework bridges these two extremes—offering parameter efficiency and fine-grained specialization within the same architecture.

\paragraph{Rank as a Bridge between PS and NPS.}
Varying $r$ seamlessly interpolates between pure parameter sharing ($r=0$) and fully distinct policies. As $r$ grows, each agent’s policy deviates more from the shared backbone, capturing complex role-specific behaviors without wholly duplicating the network. We show empirically that moderate $r$ values are sufficient for notable performance gains while retaining a low overhead in memory and computation.

\subsection{Computational and Memory Efficiency}
\label{subsec:computational_memory_efficiency}

During pretraining, LoRASA behaves like conventional PS, incurring no extra cost. Once fine-tuning begins, each agent introduces $\sum_{\ell} r \bigl(d_\ell + k_\ell\bigr)$ additional parameters, far fewer than duplicating entire networks (as in NPS). At inference, these adapters can be merged with the shared backbone (Algorithm~\ref{alg:inference}), meaning the final memory footprint remains close to a single policy, scaled only by small, rank-dependent matrices. Figures~\ref{fig:baseline_exps} and \ref{fig:resource_requirements} show that LoRASA achieves higher performance than naive PS at a fraction of NPS’s overhead, confirming its scalability.

Furthermore, by harnessing low-rank subspace adaptation, LoRASA offers an attractive middle ground—yielding heterogeneous agent behaviors with minimal resource demands. Overall, LoRASA expands the design space for cooperative multi-agent RL, moving beyond rigidly shared or fully distinct parameters toward a more adaptive paradigm.

\section{Experimental Setup}
\label{sec:experimental_setup}

We evaluate LoRA-based multi-agent reinforcement learning (MARL) across diverse continuous and discrete environments, detailing our tasks, baselines, metrics, and key findings. Our experiments demonstrate LoRASA’s ability to bridge the gap between purely shared (PS) and fully distinct (NPS) policies, reducing resource overhead while retaining agent-specific specialization. \footnote{Code is available at: \href{https://anonymous.4open.science/r/LoRASA-0D6F}{anonymous.4open.science}.}

\subsection{Environments and Tasks}
\label{subsec:environments}

\paragraph{MAMuJoCo (Continuous Control).}
We first consider MAMuJoCo, where each agent controls specific joints of a multi-limbed robot. Actions are continuous torques, and observations include local joint information (positions, velocities, etc.) plus agent IDs. Episodes run for up to 1000 steps or until the robot becomes inactive (e.g., falls). We benchmark on Half Cheetah 2x3, Walker 3x2 and Ant 4x2 under partial observability. For Humanoid $9|8$, we follow prior work~\citep{HAPPO} in providing global states to avoid degenerate solutions under severe partial observability. These tasks vary significantly in coordination needs, aiming to test LoRASA’s generality in continuous multi-agent control.

\paragraph{SMAC (Discrete Combat).}
For discrete actions, we utilize the StarCraft Multi-Agent Challenge (SMAC)~\citep{SMAC}. Unlike SMAC-v2~\citep{SMACv2}, which randomly samples agents across episodes, SMAC maintains consistent agent assignments. This consistency is crucial for training agent-specific parameters, as random sampling in SMAC-v2 could lead to some agents being trained more frequently than others, introducing unwanted complexity. In SMAC, each agent controls a StarCraft II unit with observations that include local surroundings, partial enemy and ally information, and agent IDs. We evaluate our approach on maps such as 3s5z, 1c3s5z, 3s5z\_vs\_3s6z, and MMM2. These scenarios require specialized roles—for instance, medics versus frontline marines—providing a robust test of LoRASA’s ability to learn heterogeneous behaviors without necessitating fully separate policies.

\subsection{Baselines and Comparisons}
\label{subsec:baselines}

We compare LoRASA to four baselines spanning full sharing, partial sharing, and fully separate parameters:

\begin{itemize}
    \item \textbf{PS + ID.} A single shared policy for all agents, with agent IDs appended to observations~\citep{Gupta2017CooperativeMC, CTDE_3, CTDE_4, terry2020revisiting}. Highly memory-efficient but often fails to capture diverse agent roles.
    \item \textbf{NPS.} Each agent trains an entirely separate network~\citep{HAPPO, A2PO}, allowing maximal specialization at the cost of significant resource overhead.
    \item \textbf{SePS.} Clusters agents by similarity and assigns a shared policy per cluster~\citep{christianos2021SePS}. Reduces the overhead of NPS but may perform suboptimally when agent diversity is high.
    \item \textbf{MTL.} Multi-task learning with partial sharing, typically restricting specialization to the final layer while sharing other layers between agents across tasks~\citep{MTL, MTLSurvey, MTLSurveyDL}.

\end{itemize}

LoRASA comprises two methods, \textbf{PS+LoRA} and \textbf{SePS+LoRA}, which build on top of PS and SePS, respectively. By contrast, \textbf{LoRASA} uses low-rank adaptation matrices on a shared backbone, aiming for near-NPS specialization with far lower parameter and computational demands.

\subsection{Evaluation Metrics and Protocol}
\label{subsec:evaluation_metrics}

\paragraph{Metrics.}
We measure cumulative episode return for MAMuJoCo and episodic win rate for SMAC (and episodic return see Appendix Figure~\ref{fig:smac_rewards}). We also report the total parameter count and wall-clock training and inference time, reflecting LoRASA’s resource efficiency.

\paragraph{Training Protocol.}
All methods train for up to 12 million steps, repeated over 5 random seeds for reliability. LoRA fine-tuning begins after a shared-policy pretraining phase, typically when the shared policy begins to demonstrate improved learning (see ablation studies in Figure~\ref{fig:lora_ablation}). We choose this checkpoint to balance the need for core coordination strategies (captured by the shared policy) against leaving sufficient room for agent-specific refinements. Appendix~\ref{subsec:hyperparameters} provides additional implementation details.

\subsection{Overall Performance and Resource Usage}
\label{subsec:results}

\begin{figure*}[htbp]
\centering
\begin{subfigure}
    \centering
    \includegraphics[width=\linewidth]{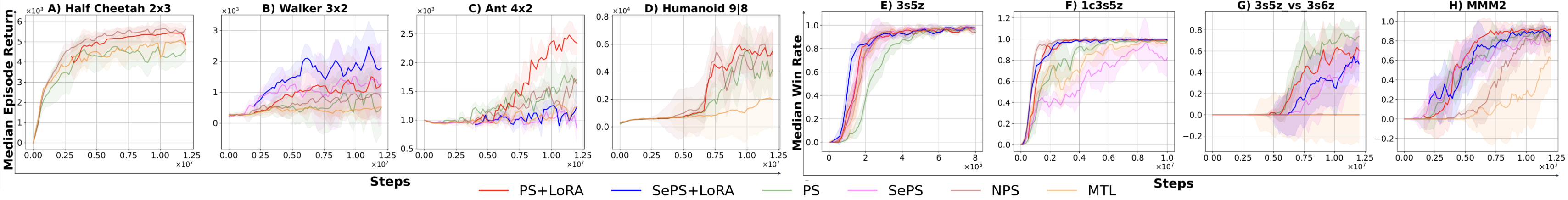}
    \label{fig:a2po_mujoco}
\end{subfigure}

\vspace{0.5em} 

\begin{subfigure}
    \centering
    \includegraphics[width=\linewidth]{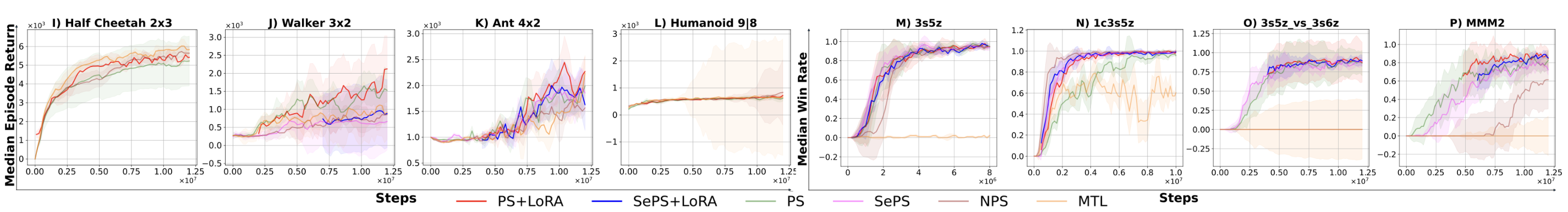}
    \label{fig:mappo_mujoco}
\end{subfigure}

\caption{Performance comparison of different parameter sharing approaches (PS, NPS, SePS, MTL, PS+LoRA and SePS+LoRA) using A2PO (row1, A--H) and MAPPO (row2, I--P) across four MAMuJoCo and SMAC scenarios: Half Cheetah 2x3, Walker 3x2, Ant 4x2, Humanoid $9|8$, 3s5z, 1c3s5z, 3s5z\_vs\_3s6z, and MMM2. The graphs plot median episode returns and evaluation win rates versus environment steps for each approach for MAMujoco and SMAC respectively. Half Cheetah 2x3 and Humanoid $9|8$ has two agents so we do not have SePS and SePS+LoRA. MAPPO learning style struggles with Humanoid $9|8$ irrespective of the parameter sharing framework.}
\label{fig:baseline_exps}
\vspace{-13pt}  
\end{figure*}

\paragraph{Performance Across Benchmarks.}
Figure~\ref{fig:baseline_exps} compares LoRASA-based approaches (\emph{PS+LoRA}, \emph{SePS+LoRA}) with the baselines on MAMuJoCo and SMAC. LoRASA frequently outperforms naive PS and, in many tasks, matches or surpasses NPS—yet at a fraction of NPS’s parameter overhead. Under A2PO, \emph{SePS+LoRA} and \emph{PS+LoRA} achieves top scores on Walker 3x2 and Ant 4x2, reflecting its ability to adapt efficiently within a shared architecture. MAPPO shows similar trends, with \emph{PS+LoRA} leading in tasks like Walker and Ant. Even in tasks where MTL and NPS performs strongly (e.g., Half Cheetah in MAPPO and A2PO respectively), LoRASA remains competitive but requires significantly less computation than NPS. Notably, all methods trained using MAPPO—including LoRASA—failed to make meaningful progress on the exceptionally challenging Humanoid $9|8$ task. This consistent struggle highlights the extreme complexity of this scenario under the MAPPO framework.

SMAC tasks show a similar pattern: LoRASA-based methods often tie or exceed the strongest baselines in scenarios like 3s5z and MMM2. Although certain maps (e.g., 3s5z\_vs\_3s6z) still favor naive PS, LoRA-based approaches remain highly effective, underscoring LoRASA’s adaptability across different MARL challenges. 

\begin{figure*}[!ht]
    \centering
    (1) \includegraphics[width=0.45\linewidth]{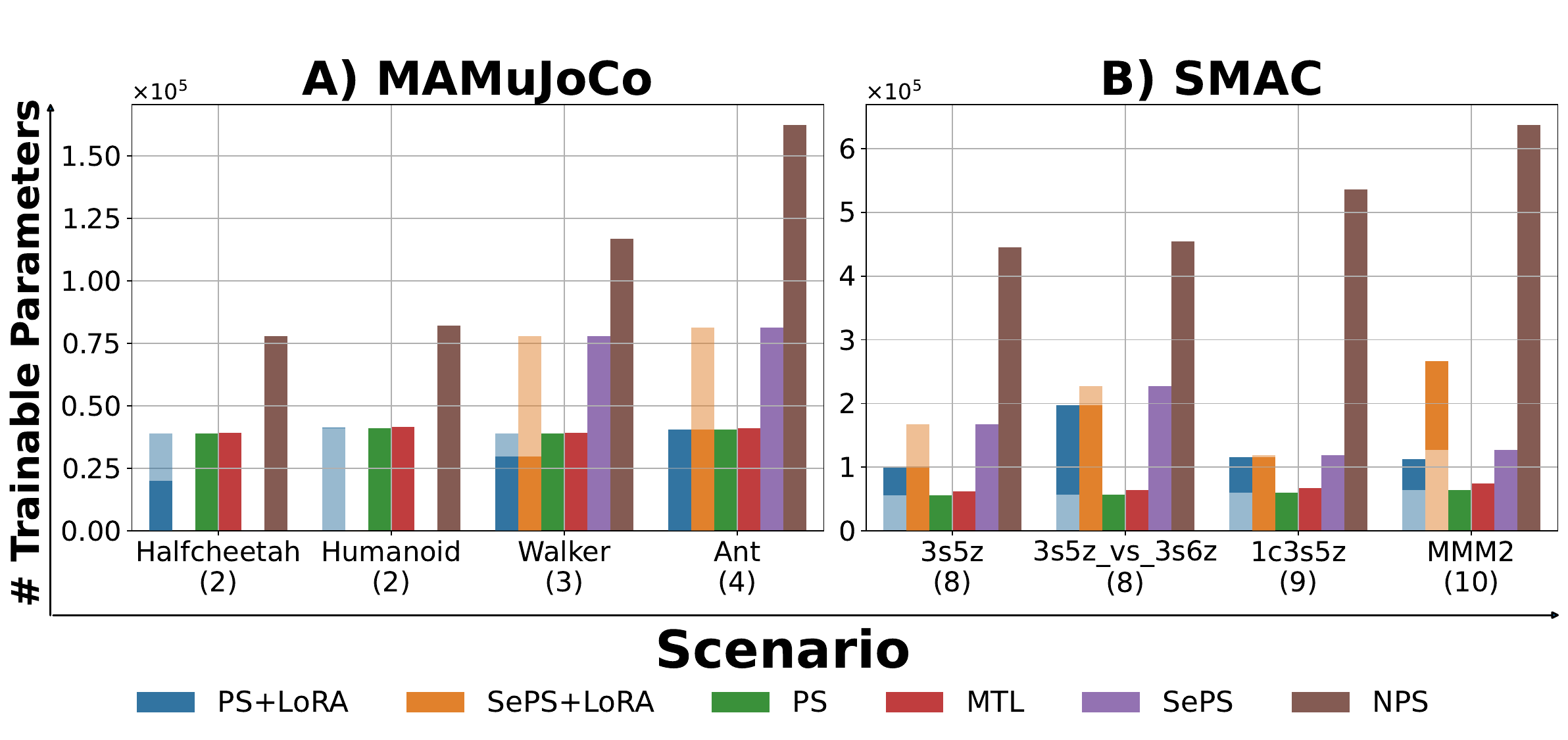}
    (2) \includegraphics[width=0.45\linewidth]{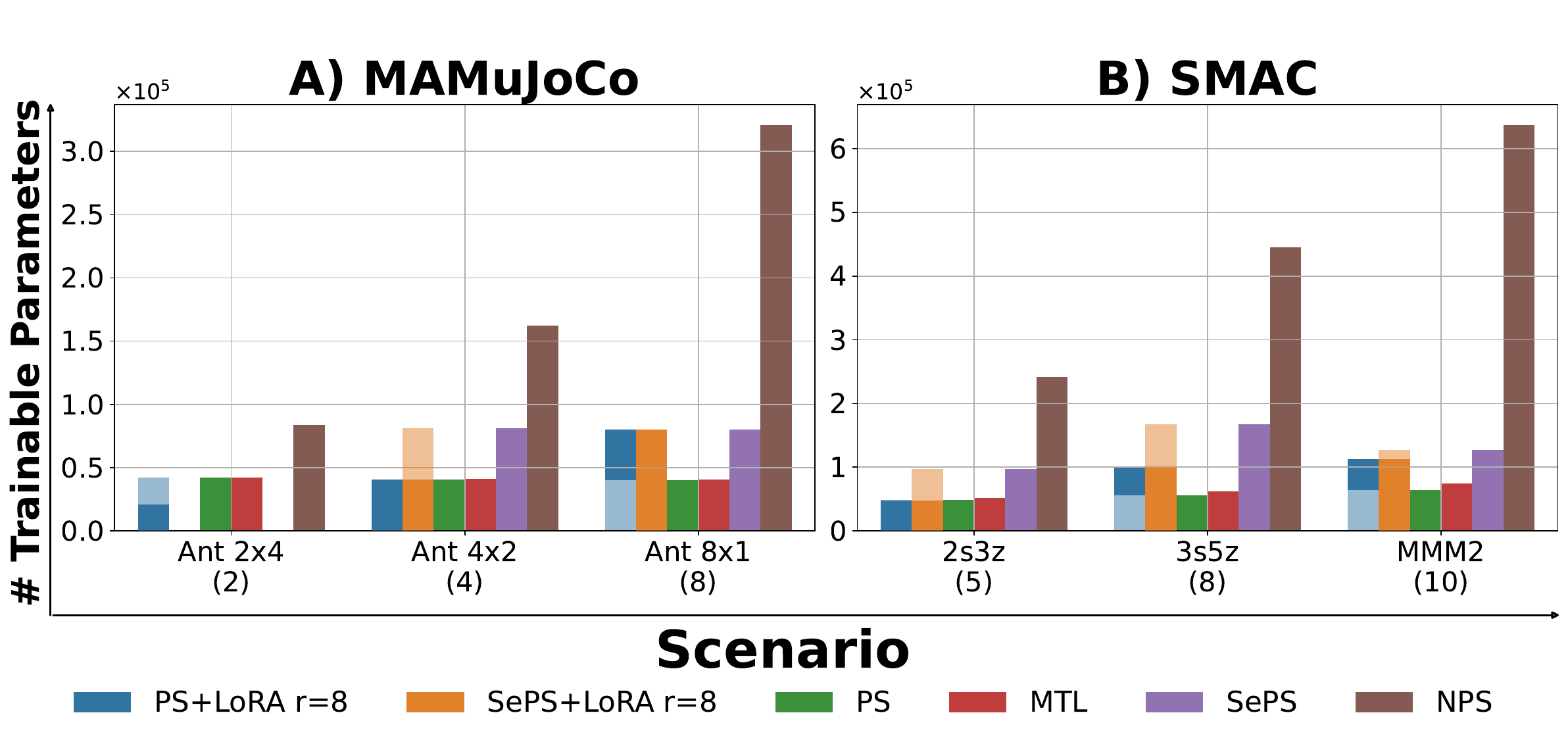}
    \\(3) \includegraphics[width=0.45\linewidth]{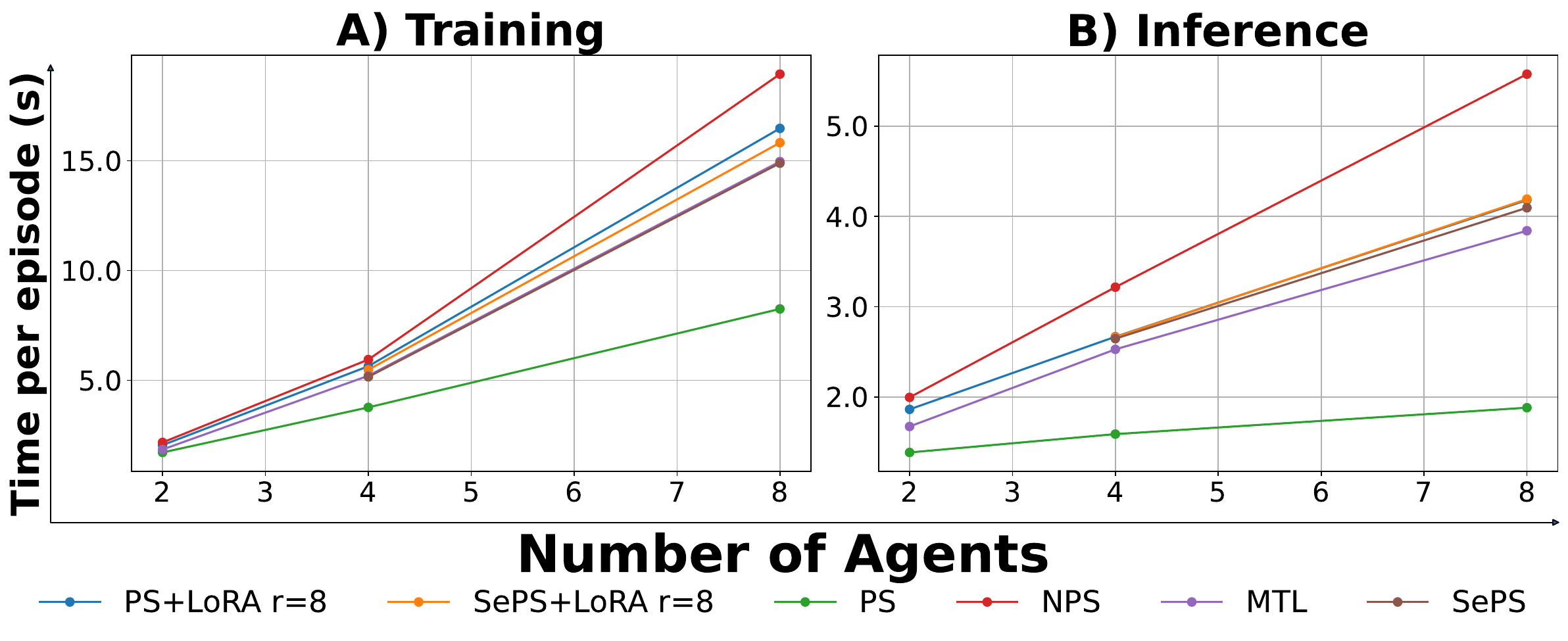}
    (4) \includegraphics[width=0.45\linewidth]{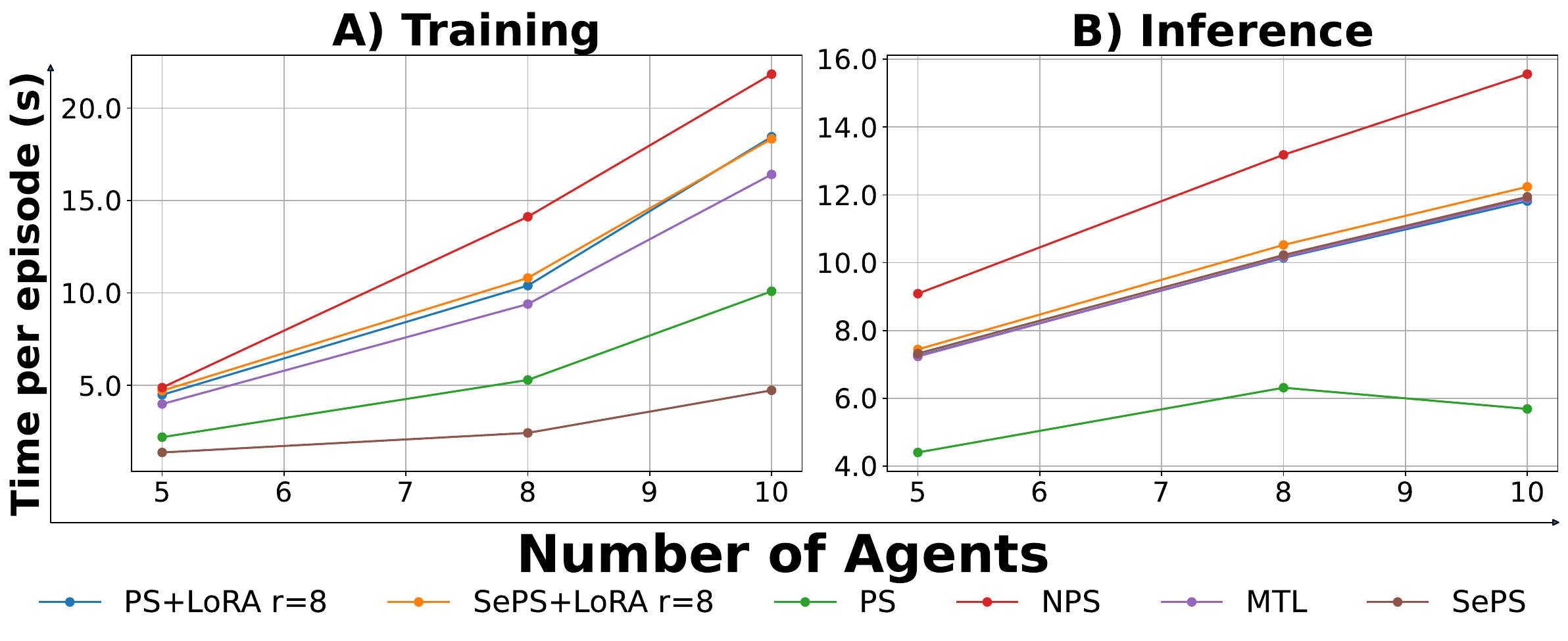}
    \caption{
        \textbf{Computational Efficiency of LoRASA Compared to Baselines.} 
        (1) \textbf{Memory footprint across environments}: Total trainable parameters for each baseline in MAMuJoCo and SMAC, highlighting LoRASA’s efficiency over NPS.
        (2) \textbf{Scalability with agent count}: Growth in trainable parameters as the number of agents increases, showing LoRASA scales efficiently while NPS grows linearly.
        (3) \textbf{Training and inference speed in MAMuJoCo}: LoRASA-based approaches significantly reduce computational time compared to NPS while achieving comparable or superior performance.
        (4) \textbf{Training and inference speed in SMAC}: Similar trends observed in SMAC, where LoRASA improves computational efficiency without compromising coordination quality.\footnotemark
    }
    \label{fig:resource_requirements}
    \vspace{-15pt}  
\end{figure*}

\paragraph{Resource Efficiency.}
Figure~\ref{fig:resource_requirements} compares parameter counts and runtime. While PS is cheapest, it often underperforms in roles requiring specialization. NPS, though powerful, scales poorly in both memory and wall-clock time. LoRASA achieves strong performance similar to (and sometimes exceeding) NPS with far fewer additional parameters. This is especially evident when scaling from 4 to 8 agents, where NPS overhead spikes but LoRASA’s cost grows moderately thanks to its low-rank updates. These findings highlight LoRASA as the “sweet spot” in MARL: it achieves the expressiveness and performance of NPS while retaining the parameter efficiency and scalability of PS, making it a practical, resource-friendly solution for large-scale MARL systems.

\footnotetext{Note: Fig (1) \& (2) Light shades of orange and blue indicate the pretraining stage (Algorithm~\ref{alg:pretraining}), while dark shades indicate fine-tuning (Algorithm~\ref{alg:finetuning}) in LoRA-based methods. For Fig (3) \& (4), MAMuJoCo (Ant 2x4, 4x2, 8x1) and SMAC (2s3z, 3s5z, MMM2)}

\subsection{Ablation Studies}
\label{subsec:ablations}

\begin{figure*}[htbp]
\centering

\includegraphics[width=0.33\linewidth]{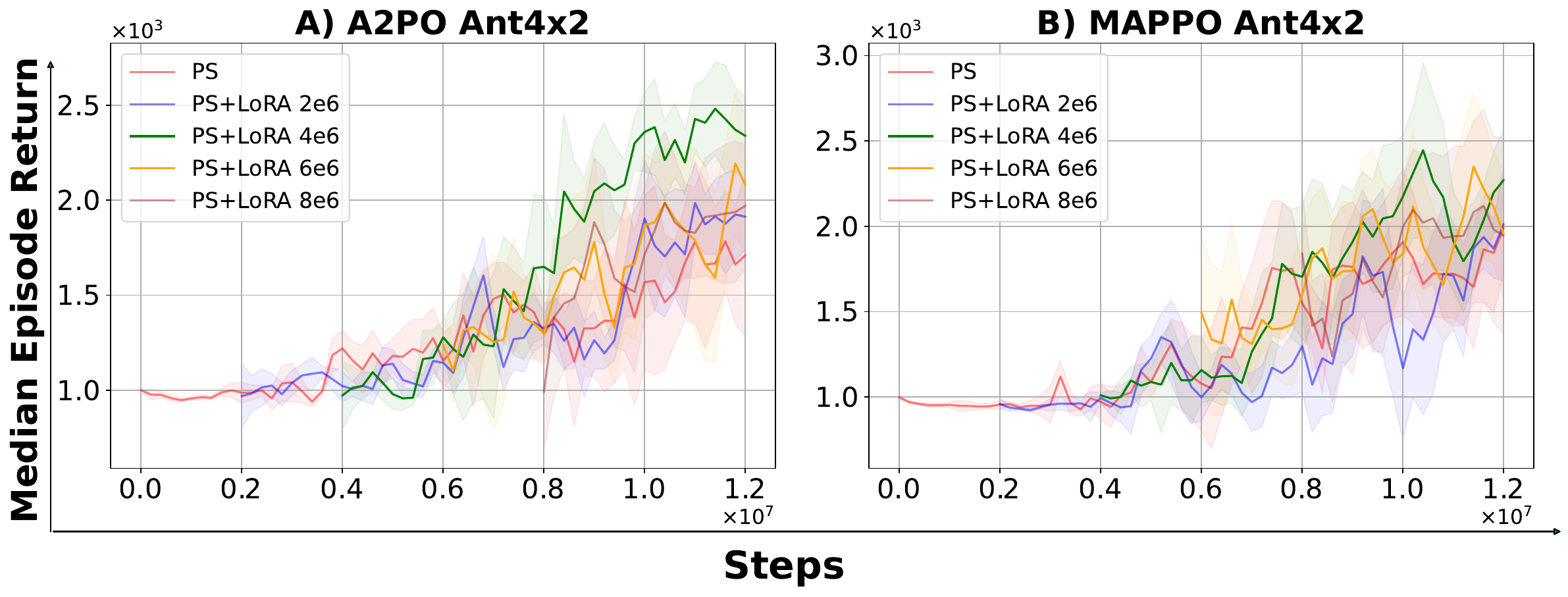}
\includegraphics[width=0.33\linewidth]{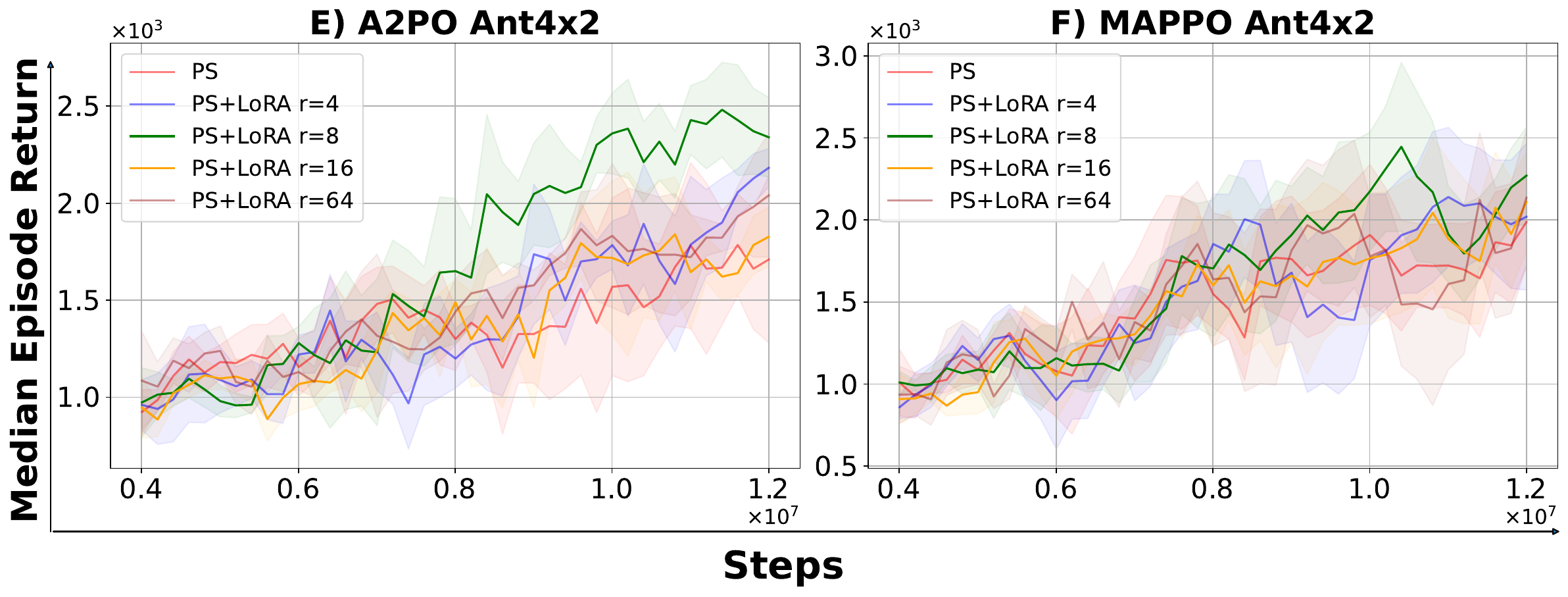}
\includegraphics[width=0.33\linewidth]{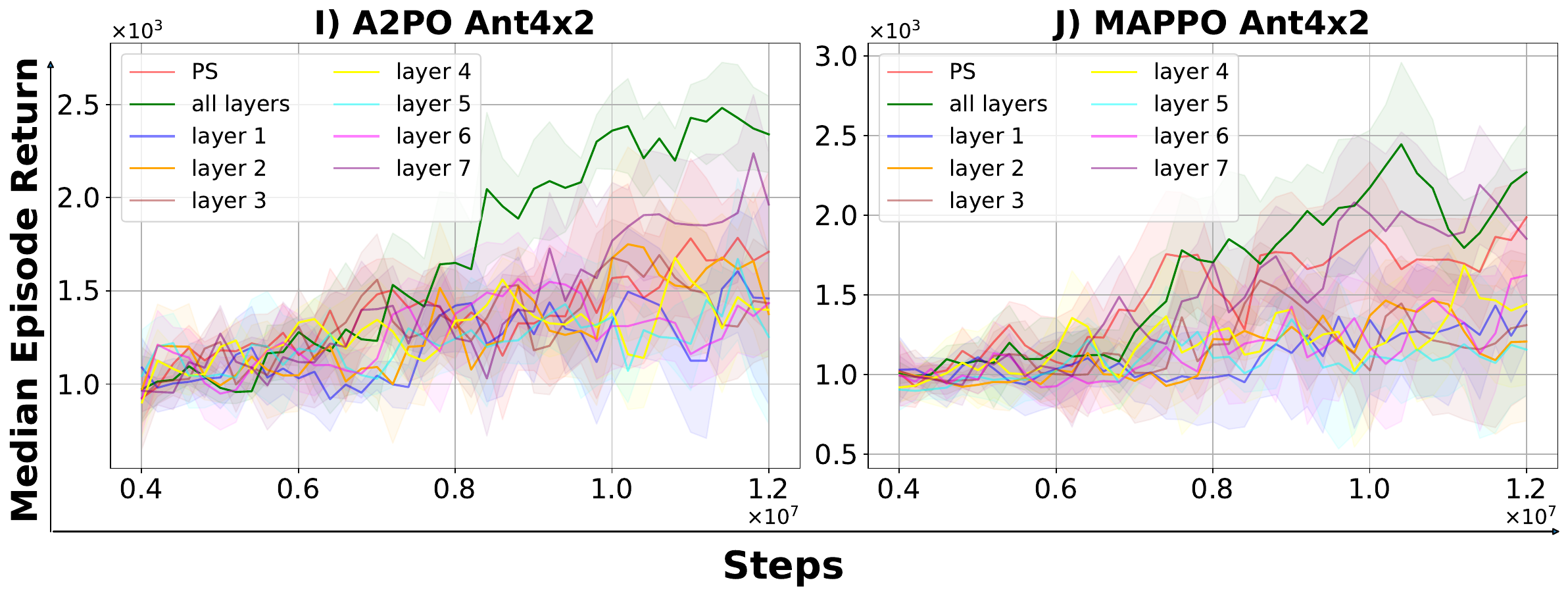}
\includegraphics[width=0.33\linewidth]{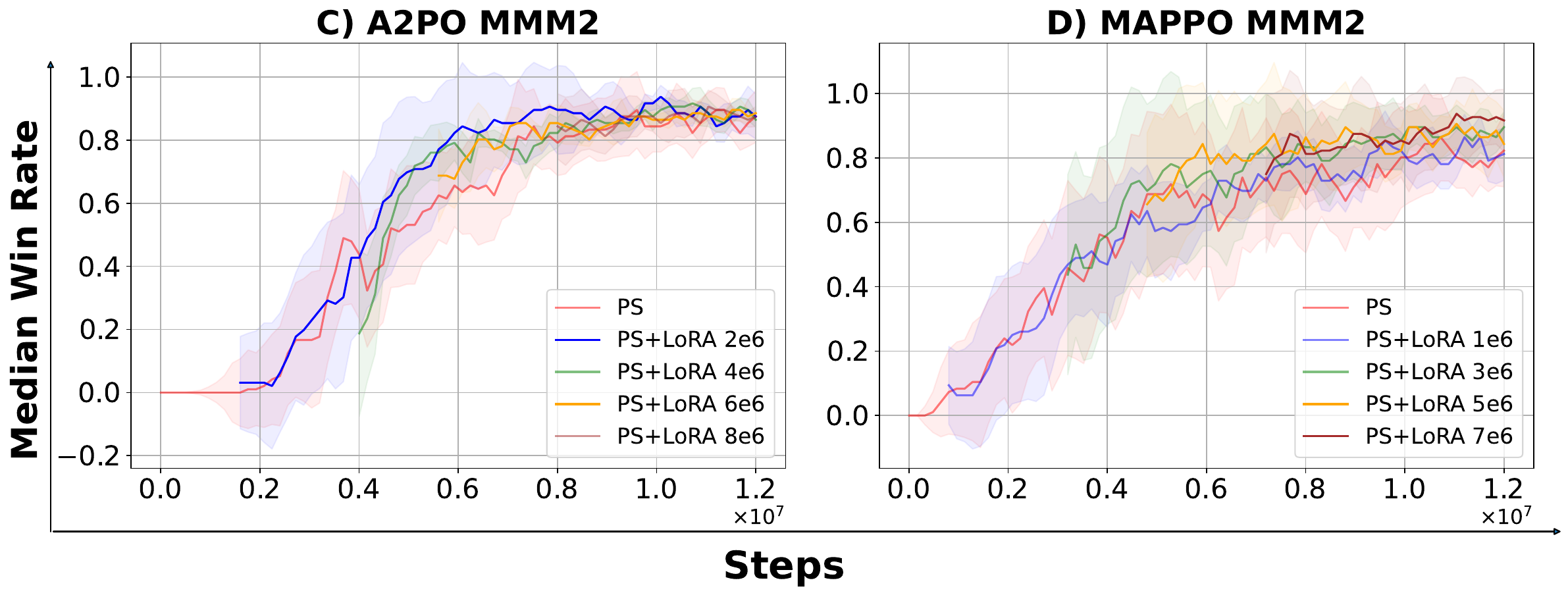}
\includegraphics[width=0.33\linewidth]{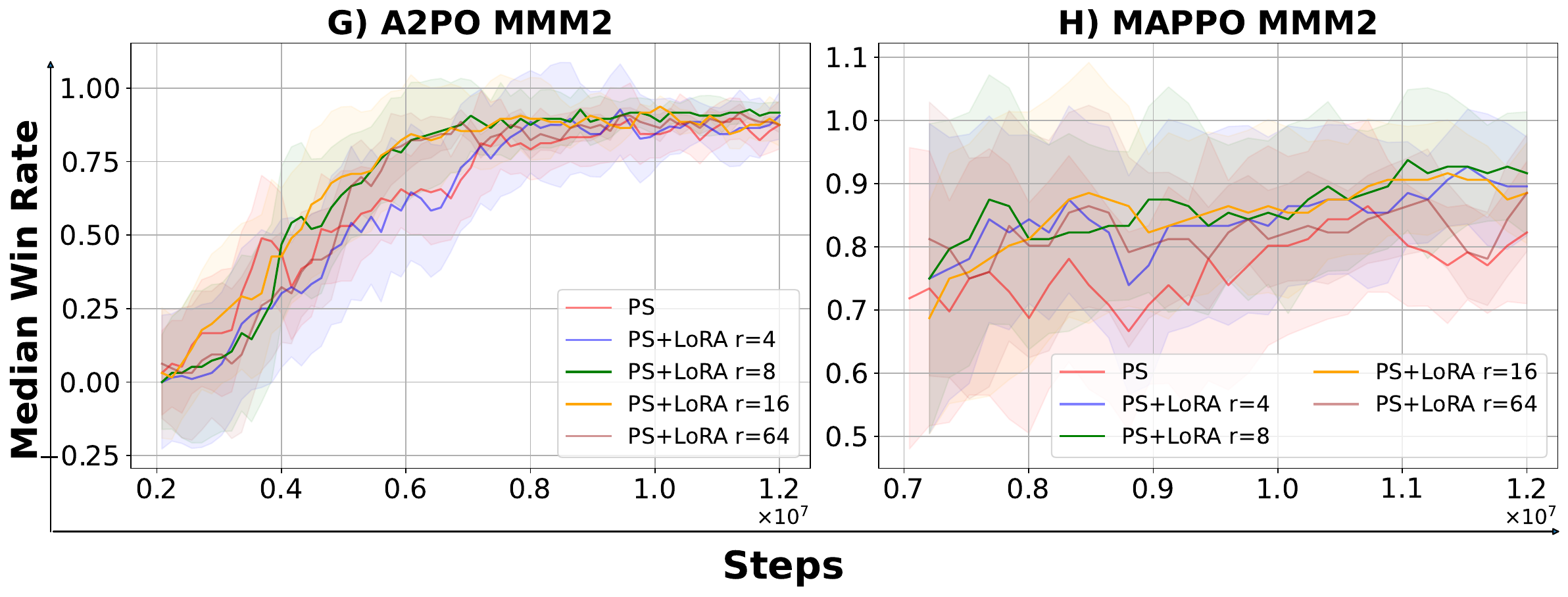}
\includegraphics[width=0.33\linewidth]{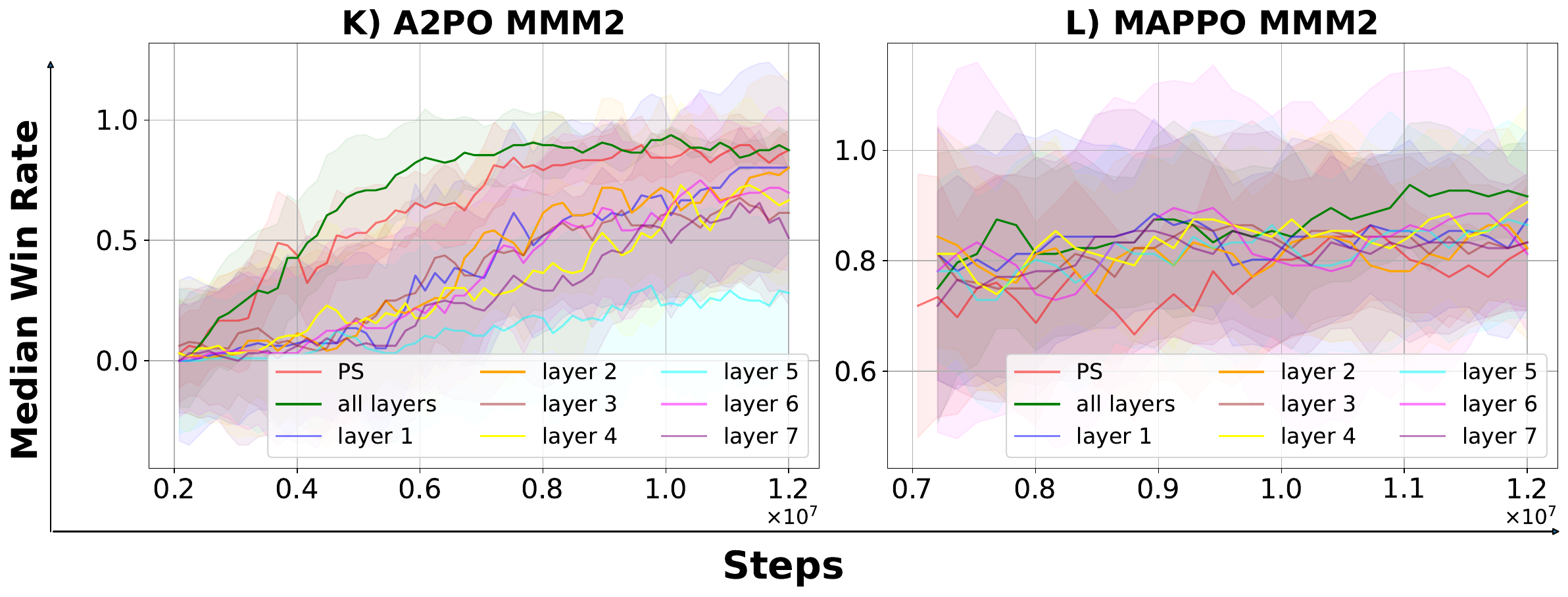}
\caption{Ablation studies on A2PO and MAPPO algorithms in Ant4x2 and MMM2 environments. (A--D) \textbf{Timing of LoRA Fine-Tuning}: Evaluates checkpoints starting at different environment steps versus the Parameter Sharing (PS) baseline. (E--H) \textbf{LoRA Rank \( r \)}: Assesses the impact of varying \( r \) values at 4, 8, 16, and 64 (full rank) compared to the PS baseline. (I--L) \textbf{Layer-Wise LoRA}: Compares the effect of applying LoRA selectively to different layers of the policy including applying LoRA to all layers simultaneously. Each subplot displays median episode returns and win rates over environment steps for MAMujoco and SMAC respectively, demonstrating LoRASA’s effectiveness in learning heterogeneous behaviors while balancing efficiency and expressivity.}

\label{fig:lora_ablation}
\vspace{-15pt}  
\end{figure*}

Our ablation experiments highlight three key dimensions—\emph{fine‐tuning checkpoints}, \emph{LoRA rank}, and \emph{adapter placement}—that underscore LoRA’s capacity to systematically unlock agent-specific policies while preserving the advantages of PS. We present ablation results for these factors using A2PO and MAPPO in two challenging scenarios, Ant 4x2 from MAMuJoCo and MMM2 from SMAC (refer Figure~\ref{fig:lora_ablation}).

\paragraph{Early vs. Late Fine-Tuning.}
Our experiments, Figs~\ref{fig:lora_ablation} (A)--(D), show that LoRA typically outperforms pure parameter sharing when introduced at checkpoints where the shared policy begins steady improvement but hasn't fully converged. For instance, switching to LoRA at \(4\times10^6\) steps works best for A2PO and MAPPO on Ant 4x2, while \(2\times10^6\) steps is optimal for A2PO on MMM2. These points likely mark the emergence of useful knowledge that LoRA can specialize more efficiently than continuing with shared weights alone.

However, in the case of MAPPO on MMM2, earlier switches only match PS performance. A later transition, around \(7\times10^6\) steps, yields peak results. This suggests that MAPPO requires more training in complex environments like MMM2 to form a robust foundation, after which LoRA can refine agent-specific behaviors without disrupting stability—consistent with the poor performance of MAPPO NPS in MMM2, see Figure~\ref{fig:baseline_exps}(P).

These findings suggest a practical guideline: \textbf{introduce LoRA updates once the shared policy exhibits competent yet non-plateaued performance}, ensuring an optimal window for effective specialization.

\paragraph{\textbf{Rank \(\mathbf{r}\): Striking a Balance Between Capacity and Efficiency.}}
We evaluate the effects of ranks $4$, $8$, $16$, and $64$ (full rank). Experiments, refer Figs~\ref{fig:lora_ablation} (E)--(H), demonstrate that moderate ranks (e.g., \(\mathbf{r=8}\)) often outperform or match full‐rank updates, reinforcing the idea that agent diversity resides in a smaller subspace than the entire parameter space. Interestingly, higher ranks (\(\mathbf{r=16}\), \(\mathbf{r=64}\)) can lead to slower convergence or overfitting, while extremely low ranks (\(\mathbf{r=4}\)) are sometimes insufficient for capturing nuanced behaviors. The sweet spot around \(\mathbf{r=8}\) suggests that LoRA’s “low‐rank” premise is more than a parameter‐efficiency hack; it’s a targeted mechanism that \emph{regularizes} agent adaptations and harnesses a smaller, behaviorally meaningful subspace. This subspace is powerful enough to drive strong performance and sample efficiency without requiring the overhead of NPS.

\paragraph{\textbf{Adapter Placement: All Layers vs. Specific Layers.}}  
Figs~\ref{fig:lora_ablation} (I)--(L) reveals that distributing LoRA across \emph{multiple} network layers generally performs best, especially when including higher/intermediate layers. By contrast, only adapting the final (output) layer sees strong but not top‐tier performance, indicating that decisions made at earlier layers are also relevant for role differentiation. Meanwhile, the minimal impact of the earliest layers suggests that certain low‐level feature extractions are already well handled by the shared backbone. LoRA’s capacity to adapt \emph{any} layer—rather than just the output or action layer—provides more robust, fine‐grained agent specialization. This finding runs counter to simpler multi‐task learning methods that adapt only the last few layers, underscoring the unique advantage of fully distributing LoRA across relevant modules.

Taken together, these ablation results show that LoRA provides a structured path to specialization while preserving the collaborative benefits of shared policy training. Full-layer LoRA generally matches or outperforms its baseline, except when applied too early or with overly low ranks. For optimal performance, we recommend introducing LoRA at a checkpoint where the shared policy shows steady improvement, using a moderate rank (\(r \approx 8\)), and adapting all layers. In more homogeneous scenarios, applying LoRA after baseline convergence can also be effective. By strategically choosing when (\(t\)) and where (\(\ell\)) to apply LoRA with a moderate rank (\(r\)), it enables near-independent policies while remaining parameter-efficient and simpler to train than fully separate networks.

\subsection{Discussion, Limitations and Future Work}
\label{subsec:discussion_limitations_future_work}

LoRASA significantly boosts performance and efficiency but comes with caveats. It relies on careful hyperparameter tuning—particularly rank selection and fine-tuning checkpoints—and depends on a robust pretrained shared policy. Fixed ranks across layers and agents may not capture the full diversity of highly heterogeneous or dynamic tasks. Moreover, LoRASA currently focuses on actor networks, leaving biases, normalization layers, and critic components unadapted. However, this approach can be extended to value-based methods; for instance, applying LoRASA to the utility functions in QMix enables agent-specific adaptations in value estimation, facilitating more nuanced coordination without requiring entirely separate policies. Future work can address these limitations by exploring dynamic rank adaptation per layer and agent type~\citep{AdaLoRA, DyLoRA}, and by extending LoRA to additional network components such as biases, normalization layers, and critic networks. Specifically, AdaLoRA~\citep{AdaLoRA}  decomposes \(\delta W\) using singular value decomposition to dynamically reduce the rank by removing less important singular values, while DyLoRA~\citep{DyLoRA} introduces adaptive mechanisms that train LoRA blocks across a range of ranks. Additionally, investigating alternating updates between shared and LoRA parameters, and integrating LoRA with hierarchical or adversarial policy architectures, could further generalize low-rank specialization. Extending this framework to competitive and adversarial multi-agent systems remains a promising direction, potentially enabling effective specialization in non-cooperative settings. These avenues promise to enhance LoRASA’s adaptability and robustness in large-scale, complex MARL applications. In challenging scenarios where the underlying algorithm struggles to learn a solid foundation (e.g., MAPPO on Humanoid $9|8$, see Figure~\ref{fig:baseline_exps}(L)), LoRASA’s effectiveness diminishes.

\section{Related Work}
\label{sec:related_work}

\paragraph{Parameter sharing (PS)} is a common strategy in MARL that reduces computational complexity by using a single policy or value network for all agents~\citep{terry2020revisiting}. However, standard PS often fails in heterogeneous environments where agents require diverse behaviors. Dynamic sharing methods~\citep{AdaPS, DynParamSharing} improve adaptability by assigning specific network modules based on predefined clusters, but they increase computational overhead, depend on role assumptions, and can introduce training instability—especially when agent roles change rapidly. In dynamic sharing methods, each agent's parameter subset can be significantly smaller than other policy baselines, making it unclear whether performance gaps stem from suboptimal selection or insufficient capacity. This scale discrepancy complicates direct comparisons with other parameter sharing approaches and is thus left out of the scope of this study. Techniques like agent identification~\citep{terry2020revisiting} or regularization-based methods~\citep{li2021CDS} attempt to differentiate agents within a shared network, but often lack sufficient representational capacity or add complexity and tuning burdens. In contrast, our approach embeds a \emph{low-rank structure} directly into shared parameters, inducing sparsity and enabling agent-specific specialization without requiring dynamic reconfiguration, clustering, or heavy regularization.

\paragraph{Selective Experience Sharing.}
Selective experience-sharing methods improve data efficiency by exchanging only relevant trajectories~\citep{SelectiveSharingExp, SharedExpAC}, reducing communication overhead and accelerating learning. However, they do not address policy expressiveness, as agents may still be constrained by a single shared model. In contrast, LoRASA operates at the parameter level, ensuring that even with fully shared transitions, low-rank offsets allow agents to develop specialized behaviors in an \(r\)-dimensional subspace. Thus, experience sharing enhances sample efficiency, while LoRASA enables representational flexibility, making them complementary rather than conflicting approaches.

\paragraph{Network Pruning} techniques~\citep{PSNetworkPruning} sparsify shared networks to lower resource usage. However, removing parameters outright risks discarding critical features needed by certain agents, especially in tasks requiring rare but crucial skills. Our work takes the opposite route: we \emph{add} low-rank modules to a shared backbone, preserving the base network and preventing irreversible performance degradation from over-pruning. This approach naturally balances expressiveness and efficiency by localizing agent-specific adaptations in small, learnable subspaces.

\paragraph{Non-Parameter Sharing (NPS)} policies (e.g., HAPPO~\citep{HAPPO}, A2PO~\citep{A2PO}) allow maximal specialization, but scale poorly in agent-heavy systems due to their linear growth in parameters and slower training due to re-learning of common behaviors. Despite their strong performance, these methods are often untenable for large populations of agents. In contrast, our low-rank approach approximates the benefits of NPS—i.e., agent-specific customization—while retaining the resource efficiency of a shared framework.

\paragraph{MARL as Multi-Task RL} methods, like ~\citep{wang2023multitaskmultiagentsharedlayers, omidshafiei2017deepdecentralizedmultitaskmultiagent, MultiTaskMARL, zhang2024hybridtrainingenhancedmultitask, PrioritizedTasKMining}, often aim to transfer knowledge such as shared decision-making modules, task representations, or agent-interactions across distinct tasks rather than to accommodate diverse roles within a single shared task. This makes them less suited for MARL scenarios where agents differ significantly but still collaborate on one global objective. In contrast, our work explicitly treats each agent as a separate “task”, applying parameter-efficient fine-tuning via low-rank adapters. Unlike approaches that only adapt output layers~\citep{MTL}, we modify internal layers as needed to capture nuanced agent behaviors without incurring the high overhead of duplicating entire networks.

\section{Conclusion}
\label{sec:conclusion}

We introduce \textbf{LoRASA}, a novel approach in MARL that integrates LoRA into parameter-sharing frameworks. LoRA enables scalable, specialized policies by constraining agent-specific updates to low-dimensional subspaces, effectively combining the efficiency of shared backbones with the expressiveness of near-independent policies. Our extensive experiments on \textbf{MAMuJoCo} and \textbf{SMAC} benchmarks demonstrate that LoRA-based methods consistently outperform or match specialized baselines like NPS, while significantly reducing both parameter and computational overhead. \textbf{Ablation studies} reveal that (1) \textbf{Deeper Network Layers} are essential for performance gains. (2) \textbf{Optimal Transition Timing} occurs when LoRA fine-tuning begins once the shared policy achieves competent, non-plateaued performance. (3) A \textbf{LoRA Rank of 8} effectively balances capacity and efficiency, scaling appropriately with task complexity. These findings provide practical guidelines for integrating LoRA into MARL pipelines. Future work will explore \textbf{dynamic rank adaptation} per layer and agent type, \textbf{alternating updates} between shared and LoRA parameters, and extending LoRA to \textbf{critic networks} and \textbf{adversarial multi-agent systems}, thereby enhancing adaptability and robustness in complex MARL environments.

\bibliography{main}

\begin{thebibliography}{45}
\providecommand{\natexlab}[1]{#1}
\providecommand{\url}[1]{\texttt{#1}}
\expandafter\ifx\csname urlstyle\endcsname\relax
  \providecommand{\doi}[1]{doi: #1}\else
  \providecommand{\doi}{doi: \begingroup \urlstyle{rm}\Url}\fi

\bibitem[Amato(2024)]{CTDE_1}
Amato, C.
\newblock An introduction to centralized training for decentralized execution in cooperative multi-agent reinforcement learning, 2024.
\newblock URL \url{https://arxiv.org/abs/2409.03052}.

\bibitem[Boutilier(1996)]{Boutilier}
Boutilier, C.
\newblock Planning, learning and coordination in multiagent decision processes.
\newblock In \emph{Proceedings of the 6th Conference on Theoretical Aspects of Rationality and Knowledge}, TARK '96, pp.\  195–210, San Francisco, CA, USA, 1996. Morgan Kaufmann Publishers Inc.
\newblock ISBN 1558604179.

\bibitem[Caruana(1997)]{MTL}
Caruana, R.
\newblock Multitask learning.
\newblock \emph{Machine Learning}, 28, 07 1997.
\newblock \doi{10.1023/A:1007379606734}.

\bibitem[Cho et~al.(2014)Cho, Van~Merri{\"e}nboer, Gulcehre, Bahdanau, Bougares, Schwenk, and Bengio]{GRU}
Cho, K., Van~Merri{\"e}nboer, B., Gulcehre, C., Bahdanau, D., Bougares, F., Schwenk, H., and Bengio, Y.
\newblock Learning phrase representations using rnn encoder-decoder for statistical machine translation.
\newblock \emph{arXiv preprint arXiv:1406.1078}, 2014.

\bibitem[Christianos et~al.(2021{\natexlab{a}})Christianos, Papoudakis, Rahman, and Albrecht]{christianos2021SePS}
Christianos, F., Papoudakis, G., Rahman, M.~A., and Albrecht, S.~V.
\newblock Scaling multi-agent reinforcement learning with selective parameter sharing.
\newblock In \emph{International Conference on Machine Learning}, pp.\  1989--1998. PMLR, 2021{\natexlab{a}}.

\bibitem[Christianos et~al.(2021{\natexlab{b}})Christianos, Schäfer, and Albrecht]{SharedExpAC}
Christianos, F., Schäfer, L., and Albrecht, S.~V.
\newblock Shared experience actor-critic for multi-agent reinforcement learning, 2021{\natexlab{b}}.
\newblock URL \url{https://arxiv.org/abs/2006.07169}.

\bibitem[Chu \& Ye(2017)Chu and Ye]{Chu2017PSDPG}
Chu, X. and Ye, H.
\newblock Parameter sharing deep deterministic policy gradient for cooperative multi-agent reinforcement learning.
\newblock \emph{CoRR}, abs/1710.00336, 2017.
\newblock URL \url{http://arxiv.org/abs/1710.00336}.

\bibitem[Crawshaw(2020)]{MTLSurveyDL}
Crawshaw, M.
\newblock Multi-task learning with deep neural networks: {A} survey.
\newblock \emph{CoRR}, abs/2009.09796, 2020.
\newblock URL \url{https://arxiv.org/abs/2009.09796}.

\bibitem[Eckart \& Young(1936)Eckart and Young]{Eckart1936TheAO}
Eckart, C. and Young, G.~M.
\newblock The approximation of one matrix by another of lower rank.
\newblock \emph{Psychometrika}, 1:\penalty0 211--218, 1936.
\newblock URL \url{https://api.semanticscholar.org/CorpusID:10163399}.

\bibitem[Ellis et~al.(2023)Ellis, Cook, Moalla, Samvelyan, Sun, Mahajan, Foerster, and Whiteson]{SMACv2}
Ellis, B., Cook, J., Moalla, S., Samvelyan, M., Sun, M., Mahajan, A., Foerster, J.~N., and Whiteson, S.
\newblock Smacv2: An improved benchmark for cooperative multi-agent reinforcement learning, 2023.
\newblock URL \url{https://arxiv.org/abs/2212.07489}.

\bibitem[Foerster et~al.(2017)Foerster, Farquhar, Afouras, Nardelli, and Whiteson]{CTDE_4}
Foerster, J., Farquhar, G., Afouras, T., Nardelli, N., and Whiteson, S.
\newblock Counterfactual multi-agent policy gradients, 2017.
\newblock URL \url{https://arxiv.org/abs/1705.08926}.

\bibitem[Gerstgrasser et~al.(2024)Gerstgrasser, Danino, and Keren]{SelectiveSharingExp}
Gerstgrasser, M., Danino, T., and Keren, S.
\newblock Selectively sharing experiences improves multi-agent reinforcement learning, 2024.
\newblock URL \url{https://arxiv.org/abs/2311.00865}.

\bibitem[Gupta et~al.(2017)Gupta, Egorov, and Kochenderfer]{Gupta2017CooperativeMC}
Gupta, J.~K., Egorov, M., and Kochenderfer, M.~J.
\newblock Cooperative multi-agent control using deep reinforcement learning.
\newblock In \emph{AAMAS Workshops}, 2017.
\newblock URL \url{https://api.semanticscholar.org/CorpusID:9421360}.

\bibitem[Haarnoja et~al.(2018)Haarnoja, Zhou, Abbeel, and Levine]{SAC}
Haarnoja, T., Zhou, A., Abbeel, P., and Levine, S.
\newblock Soft actor-critic: Off-policy maximum entropy deep reinforcement learning with a stochastic actor, 2018.
\newblock URL \url{https://arxiv.org/abs/1801.01290}.

\bibitem[Hiriart-Urruty \& Le(2013)Hiriart-Urruty and Le]{Hiriart-Urruty_Le_2013}
Hiriart-Urruty, J.-B. and Le, H.~Y.
\newblock From eckart and young approximation to moreau envelopes andvice versa.
\newblock \emph{RAIRO - Operations Research}, 47\penalty0 (3):\penalty0 299–310, 2013.
\newblock \doi{10.1051/ro/2013040}.

\bibitem[Hochreiter \& Schmidhuber(1997)Hochreiter and Schmidhuber]{LSTM}
Hochreiter, S. and Schmidhuber, J.
\newblock Long short-term memory.
\newblock \emph{Neural computation}, 9\penalty0 (8):\penalty0 1735--1780, 1997.

\bibitem[Hu et~al.(2021)Hu, Shen, Wallis, Allen-Zhu, Li, Wang, Wang, and Chen]{hu2021lora}
Hu, E.~J., Shen, Y., Wallis, P., Allen-Zhu, Z., Li, Y., Wang, S., Wang, L., and Chen, W.
\newblock Lora: Low-rank adaptation of large language models.
\newblock \emph{arXiv preprint arXiv:2106.09685}, 2021.

\bibitem[Kim \& Sung(2023)Kim and Sung]{PSNetworkPruning}
Kim, W. and Sung, Y.
\newblock Parameter sharing with network pruning for scalable multi-agent deep reinforcement learning, 2023.
\newblock URL \url{https://arxiv.org/abs/2303.00912}.

\bibitem[Kuba et~al.(2021)Kuba, Chen, Wen, Wen, Sun, Wang, and Yang]{HAPPO}
Kuba, J.~G., Chen, R., Wen, M., Wen, Y., Sun, F., Wang, J., and Yang, Y.
\newblock Trust region policy optimisation in multi-agent reinforcement learning.
\newblock \emph{CoRR}, abs/2109.11251, 2021.
\newblock URL \url{https://arxiv.org/abs/2109.11251}.

\bibitem[Kuhn(1953)]{Kuhn}
Kuhn, H.~W.
\newblock \emph{11. Extensive Games and the Problem of Information}, pp.\  193--216.
\newblock Princeton University Press, Princeton, 1953.
\newblock ISBN 9781400881970.
\newblock \doi{doi:10.1515/9781400881970-012}.
\newblock URL \url{https://doi.org/10.1515/9781400881970-012}.

\bibitem[Li et~al.(2021)Li, Wang, Wu, Zhao, Yang, and Zhang]{li2021CDS}
Li, C., Wang, T., Wu, C., Zhao, Q., Yang, J., and Zhang, C.
\newblock Celebrating diversity in shared multi-agent reinforcement learning.
\newblock \emph{Advances in Neural Information Processing Systems}, 34:\penalty0 3991--4002, 2021.

\bibitem[Li et~al.(2024)Li, Dong, Yang, Hu, Ding, li, and Gao]{MultiTaskMARL}
Li, C., Dong, S., Yang, S., Hu, Y., Ding, T., li, W., and Gao, Y.
\newblock Multi-task multi-agent reinforcement learning with interaction and task representations.
\newblock \emph{IEEE transactions on neural networks and learning systems}, PP, 10 2024.
\newblock \doi{10.1109/TNNLS.2024.3475216}.

\bibitem[Li et~al.(2023)Li, Lou, Zhang, Xu, and Fan]{AdaPS}
Li, D., Lou, N., Zhang, B., Xu, Z., and Fan, G.
\newblock Adaptive parameter sharing for multi-agent reinforcement learning, 2023.
\newblock URL \url{https://arxiv.org/abs/2312.09009}.

\bibitem[Lowe et~al.(2020)Lowe, Wu, Tamar, Harb, Abbeel, and Mordatch]{CTDE_2}
Lowe, R., Wu, Y., Tamar, A., Harb, J., Abbeel, P., and Mordatch, I.
\newblock Multi-agent actor-critic for mixed cooperative-competitive environments, 2020.
\newblock URL \url{https://arxiv.org/abs/1706.02275}.

\bibitem[Omidshafiei et~al.(2017)Omidshafiei, Pazis, Amato, How, and Vian]{omidshafiei2017deepdecentralizedmultitaskmultiagent}
Omidshafiei, S., Pazis, J., Amato, C., How, J.~P., and Vian, J.
\newblock Deep decentralized multi-task multi-agent reinforcement learning under partial observability, 2017.
\newblock URL \url{https://arxiv.org/abs/1703.06182}.

\bibitem[Rashid et~al.(2018)Rashid, Samvelyan, de~Witt, Farquhar, Foerster, and Whiteson]{CTDE_3}
Rashid, T., Samvelyan, M., de~Witt, C.~S., Farquhar, G., Foerster, J., and Whiteson, S.
\newblock Qmix: Monotonic value function factorisation for deep multi-agent reinforcement learning, 2018.
\newblock URL \url{https://arxiv.org/abs/1803.11485}.

\bibitem[Remman \& Lekkas(2024)Remman and Lekkas]{remman2024discoveringbehavioralmodesdeep}
Remman, S.~B. and Lekkas, A.~M.
\newblock Discovering behavioral modes in deep reinforcement learning policies using trajectory clustering in latent space, 2024.
\newblock URL \url{https://arxiv.org/abs/2402.12939}.

\bibitem[Samvelyan et~al.(2019)Samvelyan, Rashid, de~Witt, Farquhar, Nardelli, Rudner, Hung, Torr, Foerster, and Whiteson]{SMAC}
Samvelyan, M., Rashid, T., de~Witt, C.~S., Farquhar, G., Nardelli, N., Rudner, T. G.~J., Hung, C.-M., Torr, P. H.~S., Foerster, J., and Whiteson, S.
\newblock The starcraft multi-agent challenge, 2019.
\newblock URL \url{https://arxiv.org/abs/1902.04043}.

\bibitem[Schneider et~al.(2024)Schneider, Schumacher, Guist, Chen, Häufle, Schölkopf, and Büchler]{schneider2024identifyingpolicygradientsubspaces}
Schneider, J., Schumacher, P., Guist, S., Chen, L., Häufle, D., Schölkopf, B., and Büchler, D.
\newblock Identifying policy gradient subspaces, 2024.
\newblock URL \url{https://arxiv.org/abs/2401.06604}.

\bibitem[Shapley(1953)]{Shapely}
Shapley, L.~S.
\newblock Stochastic games*.
\newblock \emph{Proceedings of the National Academy of Sciences}, 39\penalty0 (10):\penalty0 1095--1100, 1953.
\newblock \doi{10.1073/pnas.39.10.1095}.
\newblock URL \url{https://www.pnas.org/doi/abs/10.1073/pnas.39.10.1095}.

\bibitem[Sun \& Fazli(2019)Sun and Fazli]{Sun2019RealtimePD}
Sun, Y. and Fazli, P.
\newblock Real-time policy distillation in deep reinforcement learning.
\newblock \emph{ArXiv}, abs/1912.12630, 2019.
\newblock URL \url{https://api.semanticscholar.org/CorpusID:209515823}.

\bibitem[Sun \& Zhang(2022)Sun and Zhang]{ensemble_policy_distillation}
Sun, Y. and Zhang, Q.
\newblock Ensemble policy distillation with reduced data distribution mismatch.
\newblock In \emph{2022 International Joint Conference on Neural Networks (IJCNN)}, pp.\  1--8, 2022.
\newblock \doi{10.1109/IJCNN55064.2022.9892503}.

\bibitem[Sunehag et~al.(2017)Sunehag, Lever, Gruslys, Czarnecki, Zambaldi, Jaderberg, Lanctot, Sonnerat, Leibo, Tuyls, and Graepel]{CTDE_5}
Sunehag, P., Lever, G., Gruslys, A., Czarnecki, W.~M., Zambaldi, V., Jaderberg, M., Lanctot, M., Sonnerat, N., Leibo, J.~Z., Tuyls, K., and Graepel, T.
\newblock Value-decomposition networks for cooperative multi-agent learning, 2017.
\newblock URL \url{https://arxiv.org/abs/1706.05296}.

\bibitem[Terry et~al.(2020)Terry, Grammel, Son, Black, and Agrawal]{terry2020revisiting}
Terry, J.~K., Grammel, N., Son, S., Black, B., and Agrawal, A.
\newblock Revisiting parameter sharing in multi-agent deep reinforcement learning.
\newblock \emph{arXiv preprint arXiv:2005.13625}, 2020.

\bibitem[Valipour et~al.(2023)Valipour, Rezagholizadeh, Kobyzev, and Ghodsi]{DyLoRA}
Valipour, M., Rezagholizadeh, M., Kobyzev, I., and Ghodsi, A.
\newblock Dylora: Parameter efficient tuning of pre-trained models using dynamic search-free low-rank adaptation, 2023.
\newblock URL \url{https://arxiv.org/abs/2210.07558}.

\bibitem[Wadhwania et~al.(2019)Wadhwania, Kim, Omidshafiei, and How]{wadhwania2019policydistillationvaluematching}
Wadhwania, S., Kim, D.-K., Omidshafiei, S., and How, J.~P.
\newblock Policy distillation and value matching in multiagent reinforcement learning, 2019.
\newblock URL \url{https://arxiv.org/abs/1903.06592}.

\bibitem[Wang et~al.(2023{\natexlab{a}})Wang, Zhao, Cao, Feng, Qin, and Yu]{wang2023multitaskmultiagentsharedlayers}
Wang, J., Zhao, J., Cao, Z., Feng, R., Qin, R., and Yu, Y.
\newblock Multi-task multi-agent shared layers are universal cognition of multi-agent coordination, 2023{\natexlab{a}}.
\newblock URL \url{https://arxiv.org/abs/2312.15674}.

\bibitem[Wang et~al.(2023{\natexlab{b}})Wang, Tian, Wan, Wen, Wang, and Zhang]{A2PO}
Wang, X., Tian, Z., Wan, Z., Wen, Y., Wang, J., and Zhang, W.
\newblock Order matters: Agent-by-agent policy optimization, 2023{\natexlab{b}}.
\newblock URL \url{https://arxiv.org/abs/2302.06205}.

\bibitem[Yu et~al.(2022)Yu, Velu, Vinitsky, Gao, Wang, Bayen, and Wu]{MAPPO}
Yu, C., Velu, A., Vinitsky, E., Gao, J., Wang, Y., Bayen, A., and Wu, Y.
\newblock The surprising effectiveness of ppo in cooperative, multi-agent games, 2022.
\newblock URL \url{https://arxiv.org/abs/2103.01955}.

\bibitem[Yu et~al.(2023)Yu, Yin, Zhang, and Huang]{PrioritizedTasKMining}
Yu, Y., Yin, Q., Zhang, J., and Huang, K.
\newblock Prioritized tasks mining for multi-task cooperative multi-agent reinforcement learning.
\newblock In \emph{Proceedings of the 2023 International Conference on Autonomous Agents and Multiagent Systems}, AAMAS '23, pp.\  1615–1623, Richland, SC, 2023. International Foundation for Autonomous Agents and Multiagent Systems.
\newblock ISBN 9781450394321.

\bibitem[Yu et~al.(2024)Yu, Yin, Zhang, Xu, and Huang]{DynParamSharing}
Yu, Y., Yin, Q., Zhang, J., Xu, P., and Huang, K.
\newblock Admn: Agent-driven modular network for dynamic parameter sharing in cooperative multi-agent reinforcement learning.
\newblock In Larson, K. (ed.), \emph{Proceedings of the Thirty-Third International Joint Conference on Artificial Intelligence, {IJCAI-24}}, pp.\  302--310. International Joint Conferences on Artificial Intelligence Organization, 8 2024.
\newblock \doi{10.24963/ijcai.2024/34}.
\newblock URL \url{https://doi.org/10.24963/ijcai.2024/34}.
\newblock Main Track.

\bibitem[Zhang et~al.(2021)Zhang, Yang, and Ba{\c{s}}ar]{zhang2021multi}
Zhang, K., Yang, Z., and Ba{\c{s}}ar, T.
\newblock Multi-agent reinforcement learning: A selective overview of theories and algorithms.
\newblock \emph{Handbook of reinforcement learning and control}, pp.\  321--384, 2021.

\bibitem[Zhang et~al.(2024)Zhang, Su, He, and Sartoretti]{zhang2024hybridtrainingenhancedmultitask}
Zhang, M., Su, S., He, C., and Sartoretti, G.
\newblock Hybrid training for enhanced multi-task generalization in multi-agent reinforcement learning, 2024.
\newblock URL \url{https://arxiv.org/abs/2408.13567}.

\bibitem[Zhang et~al.(2023)Zhang, Chen, Bukharin, Karampatziakis, He, Cheng, Chen, and Zhao]{AdaLoRA}
Zhang, Q., Chen, M., Bukharin, A., Karampatziakis, N., He, P., Cheng, Y., Chen, W., and Zhao, T.
\newblock Adalora: Adaptive budget allocation for parameter-efficient fine-tuning, 2023.
\newblock URL \url{https://arxiv.org/abs/2303.10512}.

\bibitem[Zhang \& Yang(2017)Zhang and Yang]{MTLSurvey}
Zhang, Y. and Yang, Q.
\newblock A survey on multi-task learning.
\newblock \emph{CoRR}, abs/1707.08114, 2017.
\newblock URL \url{http://arxiv.org/abs/1707.08114}.

\end{thebibliography}
\bibliographystyle{icml2025}

\pagebreak
\appendix
\onecolumn
\section{Appendix}

\subsection{Pseudocode}

\begin{algorithm}
\caption{Phase 1: Shared Policy Pretraining}
\label{alg:pretraining}
\textbf{Input:} $N$: number of agents, $\text{Env}$, $\text{Algorithm}$ (e.g., MAPPO/A2PO), $\theta_{\text{shared}}$: shared parameters, $P_\text{steps}$: pretraining steps \\
\textbf{Output:} Pretrained $\theta_{\text{shared}}$
\begin{algorithmic}[1]
\STATE Initialize $\theta_{\text{shared}}$:
\FOR{$\text{step} \gets 1$ to $P_\text{steps}$}
    \STATE Collect joint trajectories $(\text{obs},\text{actions},\text{rewards},\text{next\_obs})$ from $\text{Env}$
    \STATE $\theta_{shared} \gets \text{Algorithm}.\text{update\_shared}(\theta_{\text{shared}}, \text{trajectories})$
\ENDFOR
\OUTPUT $\theta_{\text{shared}}$
\end{algorithmic}
\end{algorithm}

\begin{algorithm}
\caption{Phase 2: LoRA-Based Fine-Tuning}
\label{alg:finetuning}
\textbf{Input:} $N$: number of agents, $\text{Env}$, $\text{Algorithm}$, pretrained $\theta_{\text{shared}}$, rank $r$, $F_\text{steps}$ \\
\textbf{Output:} Agent-specific LoRA adapters $\{A_i^\ell, B_i^\ell\}$

\begin{algorithmic}
\small 
\STATE Introduce LoRA adapters $A_i^\ell,B_i^\ell$; freeze $\theta_{\text{shared}}$
\STATE \hspace{1em} $A_i^\ell \gets \mathbf{0}_{d_\ell\times r}, \quad B_i^\ell \gets \text{Random}(k_\ell\times r)$
\FOR{$\text{step} \gets 1$ to $F_\text{steps}$}
    \STATE Collect trajectories $(\text{obs}, a, \text{rewards}, \text{next\_obs})$
    \FOR{$i \gets 1$ to $N$}
        \STATE $(A_i^\ell, B_i^\ell) \gets \text{Algorithm}.\text{update\_agent\_lora}(\theta_{\text{shared}}, A_i^\ell, B_i^\ell,$
        \STATE \hspace{2em} $\text{trajectories}[i], r)$
    \ENDFOR
\ENDFOR
\OUTPUT $\{A_i^\ell, B_i^\ell\}_{i=1}^N$
\end{algorithmic}
\end{algorithm}

\begin{algorithm}
\caption{Inference with LoRA}
\label{alg:inference}
\textbf{Input:} Pretrained $\theta_{\text{shared}}$, LoRA adapters $\{A_i^\ell, B_i^\ell\}$, agent observations $\{o_i\}_{i=1}^N$ \\
\textbf{Output:} Actions $\{a_i\}_{i=1}^N$
\begin{algorithmic}[1]
\FOR{$i \gets 1$ to $N$}
    \FOR{layer $\ell$ in actor network}
        \STATE $\theta_i^\ell \gets \theta_{\text{shared}}^\ell + A_i^\ell B_i^{\ell\top}$
    \ENDFOR
\ENDFOR
\STATE $\{a_i\}_{i=1}^N \gets \text{Actor}.\text{select\_action}(\{o_i\}_{i=1}^N, \{\theta_i^\ell\}_{i=1}^N)$
\OUTPUT $\{a_i\}_{i=1}^N$
\end{algorithmic}
\end{algorithm}

\newpage
\subsection{SMAC Evaluation Rewards}
\begin{figure}[!htbp]
    \centering

    \includegraphics[width=\linewidth]{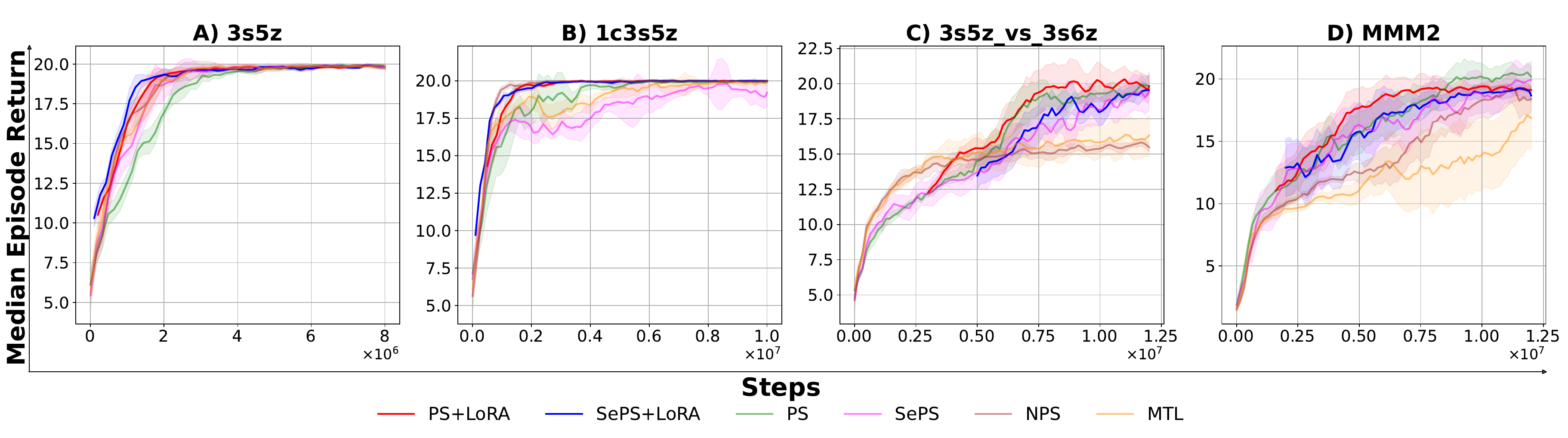} 

    \includegraphics[width=\linewidth]{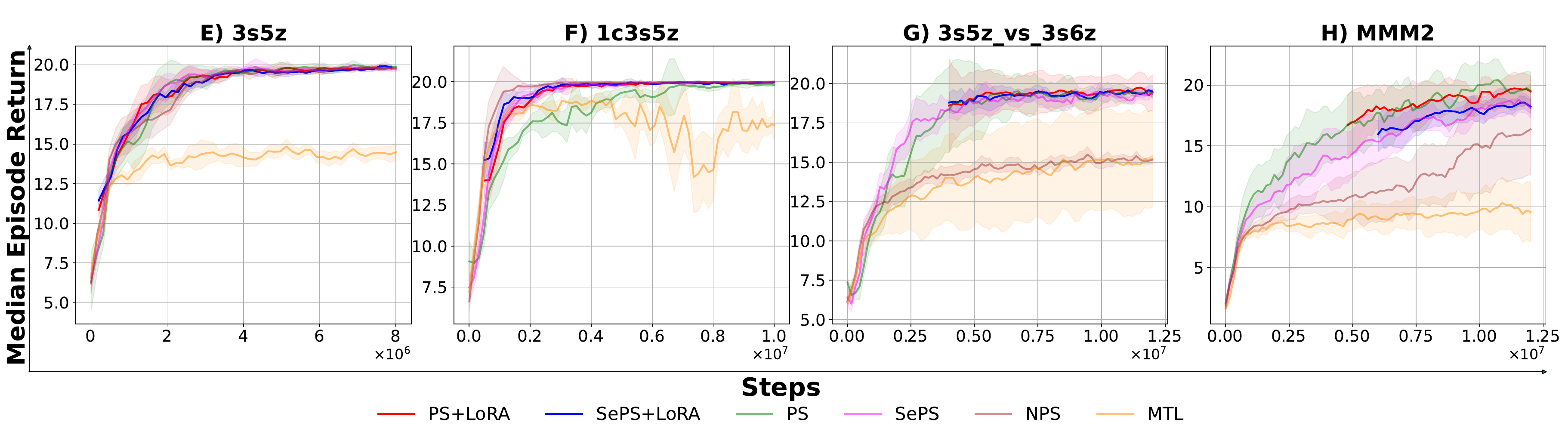}

    \caption{Evaluation episode rewards of (A--D) A2PO and (E--H) MAPPO in SMAC scenarios}
    \label{fig:smac_rewards}
\end{figure}

\clearpage

\subsection{Further Analysis}
\label{subsec:analysis}

\paragraph{Heterogeneous nature of agent policies}
\label{subsec:heterogeneous_nature} 
Inspection of the \emph{policy activation heatmaps} (see Figures ~\ref{fig:hm_a2po_3s56z} -- ~\ref{fig:hm_a2po_ant4x2_8M}) indicates that the initial layers of each agent’s policy are comparatively similar—both relative to the shared baseline and among different agents. This aligns closely with our \emph{layer ablation} findings (Figure~\ref{fig:lora_ablation}(I)--(L)), where adapting these early layers alone provides only modest returns. These early, near-identical activations suggest that fundamental state or feature extraction is largely \emph{universal}, capturing environmental signals (e.g., basic positional inputs in MAMuJoCo or unit attributes in SMAC) that all agents need in a shared manner \citep{Gupta2017CooperativeMC, terry2020revisiting}.

In contrast, later layers exhibit far more divergence in the heatmaps, signaling agent-specific computations that reflect distinct strategic roles. For example, in Figure~\ref{fig:lora_ablation}(I)--(L), applying LoRA to mid-to-high layers substantially boosts performance, underscoring that the majority of adaptive capacity is needed where the policy makes higher-level decisions (e.g., unit targeting in SMAC or joint coordination in MAMuJoCo). Further, adapting all layers emerges as the best configuration, indicating that—even though initial layers are mostly similar—some specialized nuance in lower layers can still yield incremental gains when combined with deeper-layer updates, consistent with the agent-specific differences revealed in Figures ~\ref{fig:hm_a2po_3s56z} -- ~\ref{fig:hm_a2po_ant4x2_8M}.

\begin{figure}
    \centering
    \begin{subfigure}
        \centering
        \includegraphics[width=0.4\linewidth]{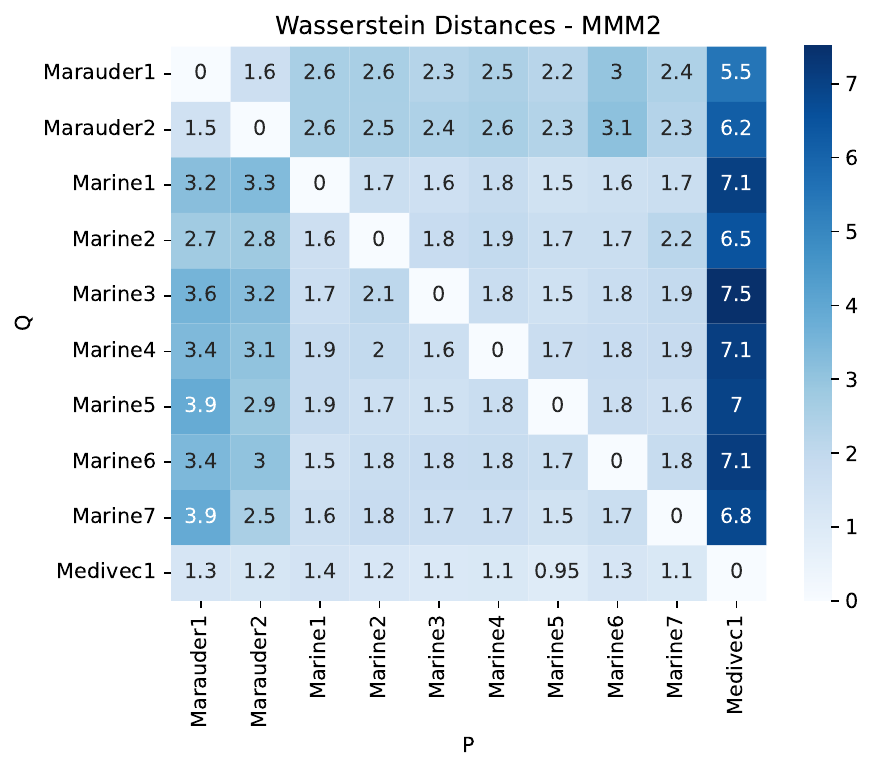} 
        \caption{Wasserstein distance MMM2}
        \label{fig:was_mmm2}
    \end{subfigure}
    \begin{subfigure}
        \centering
        \includegraphics[width=0.4\linewidth]{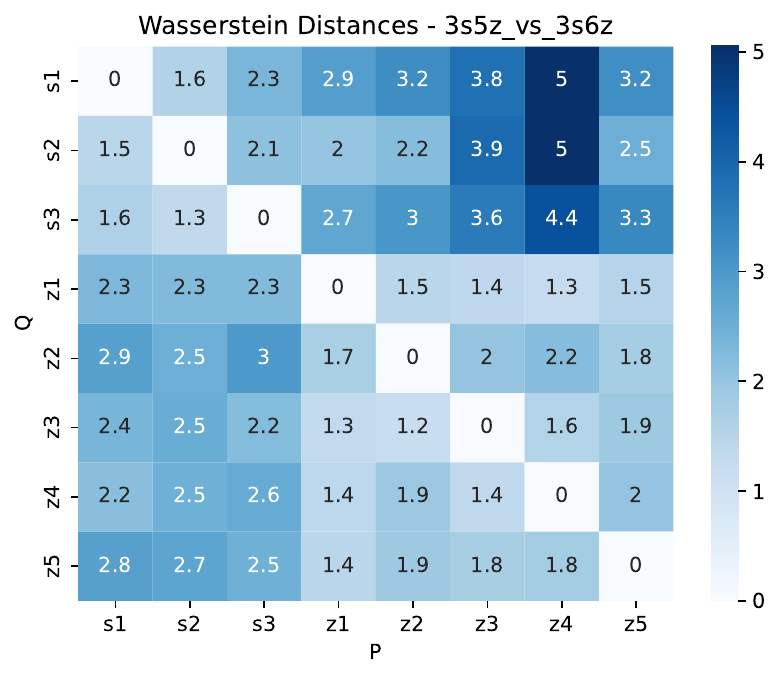} 
        \caption{Wasserstein distance 3s5z\_vs\_3s6z}
        \label{fig:was_3s5z_vs_3s6z}
    \end{subfigure}
\end{figure}

To quantify these visual distinctions, we measure the Wasserstein distance between policy distributions across agents (see Figure~\ref{fig:was_3s5z_vs_3s6z} and ~\ref{fig:was_mmm2}). Two key observations arise:

\textbf{Agents with Similar Roles but Divergent Strategies}.
Units of the same “category” often have smaller pairwise Wasserstein distances, suggesting a shared skill set or baseline. However, even among these nominally similar agents, divergences can arise—particularly in later layers—because each agent may develop a unique strategy. This behavior echoes recent work in multi-task RL that finds role similarity does not preclude agent-specific policy refinements when higher-level decisions are at play \citep{zhang2021multi, MultiTaskMARL}.

\textbf{Different Categories, Greater Distances}.
When comparing agents of distinct roles, the Wasserstein distance grows larger. This supports the notion that LoRA fosters substantial heterogeneity for roles requiring fundamentally different behaviors. Our ablation results Figure~\ref{fig:lora_ablation}(I)--(L) reinforce that focusing LoRA updates on deeper layers, where these role-specific divergences manifest, provides significant performance gains.

\paragraph{Sparsity analysis}
\label{paragraph:sparsity_analysis}

\begin{figure}
\begin{center}
\includegraphics[width=0.4\linewidth]{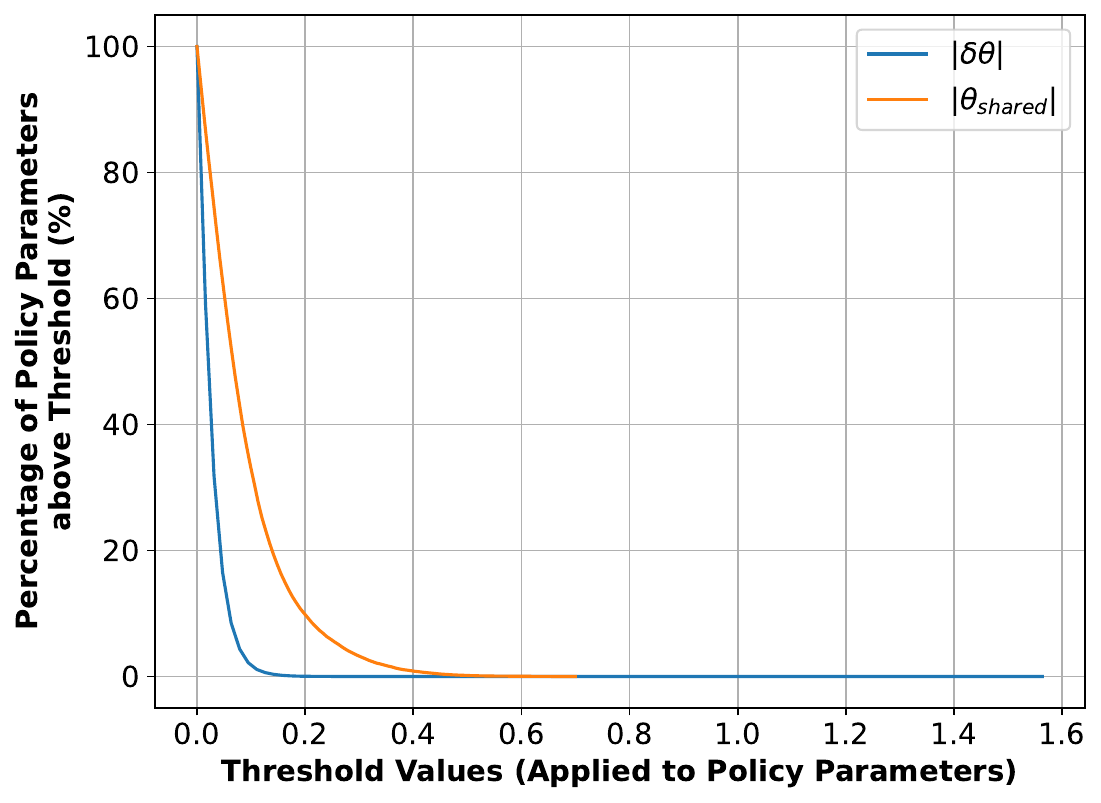}
\end{center}
\caption{Sparsity comparison between LoRA layers vs Shared layers. Percentage of policy parameters above threshold values is computed by flattening all weights into a single array, taking the absolute values, and evaluating 100 evenly spaced threshold values between the minimum and maximum of the array (inclusive). For each threshold, the percentage is calculated as \((\text{array} \geq \text{threshold}).\text{mean()} \times 100\), reflecting the proportion of parameters exceeding the given threshold.}
\label{fig:sparsity_comparison}
\end{figure}

\begin{figure}
\begin{center}
\includegraphics[width=\linewidth]{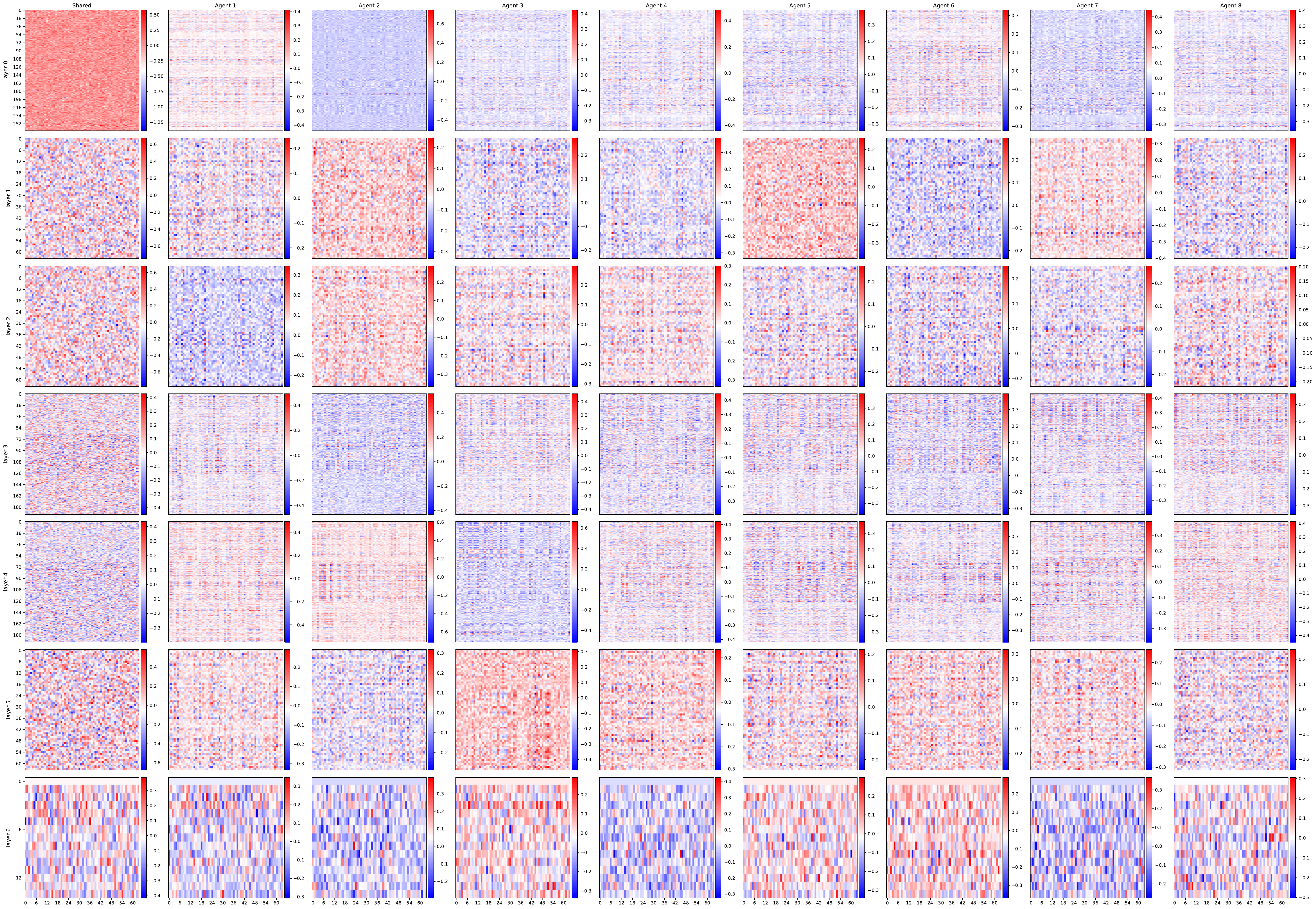}
\end{center}
\caption{Layer Activation of the map 3s5z\_vs\_3s6z using A2PO}
\label{fig:hm_a2po_3s56z}
\end{figure}

\begin{figure}
\begin{center}
\includegraphics[width=\linewidth]{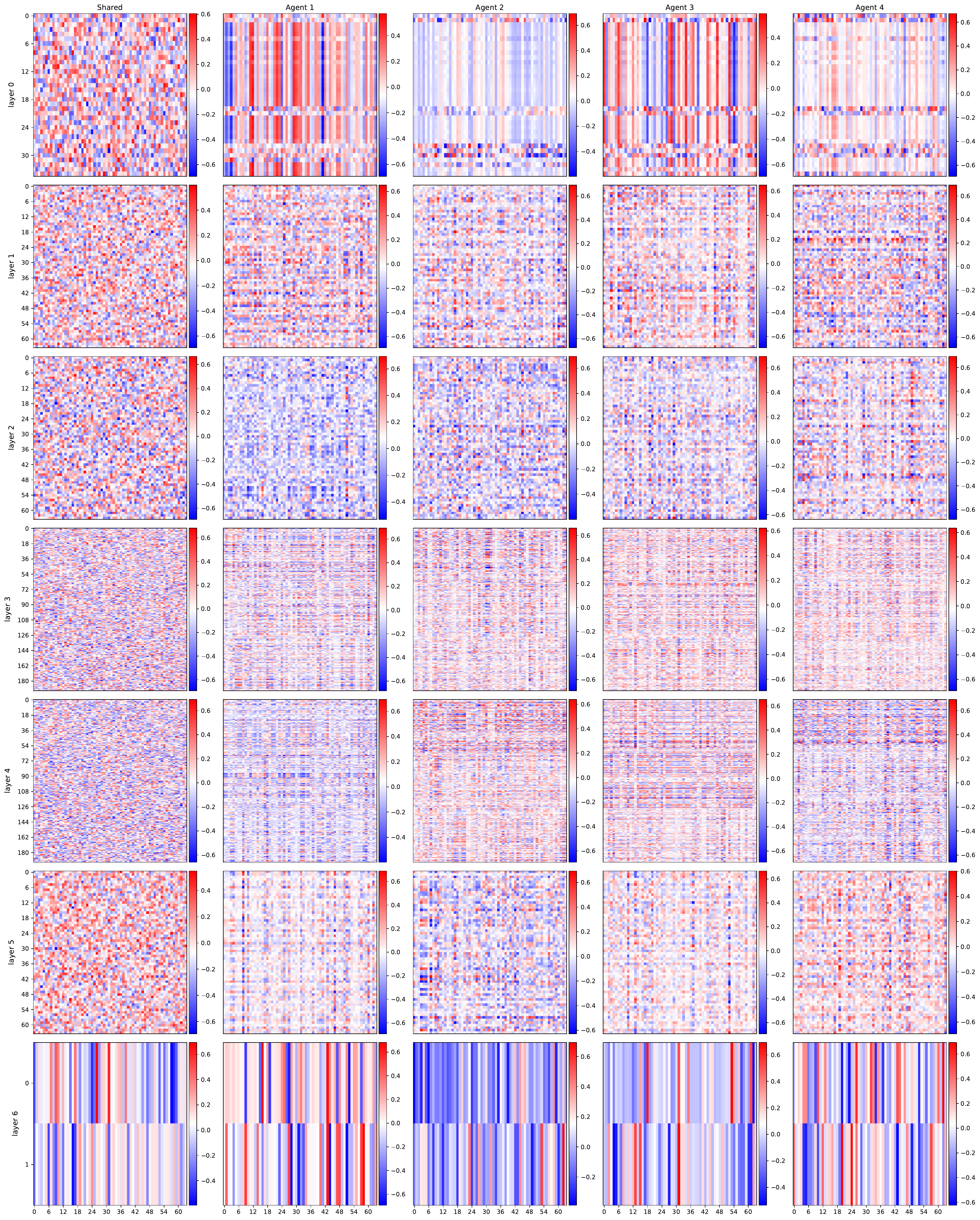}
\end{center}
\caption{Layer Activation of the map Ant4x2 using A2PO}
\label{fig:hm_a2po_ant4x2}
\end{figure}

\begin{figure}
\begin{center}
\includegraphics[width=\linewidth]{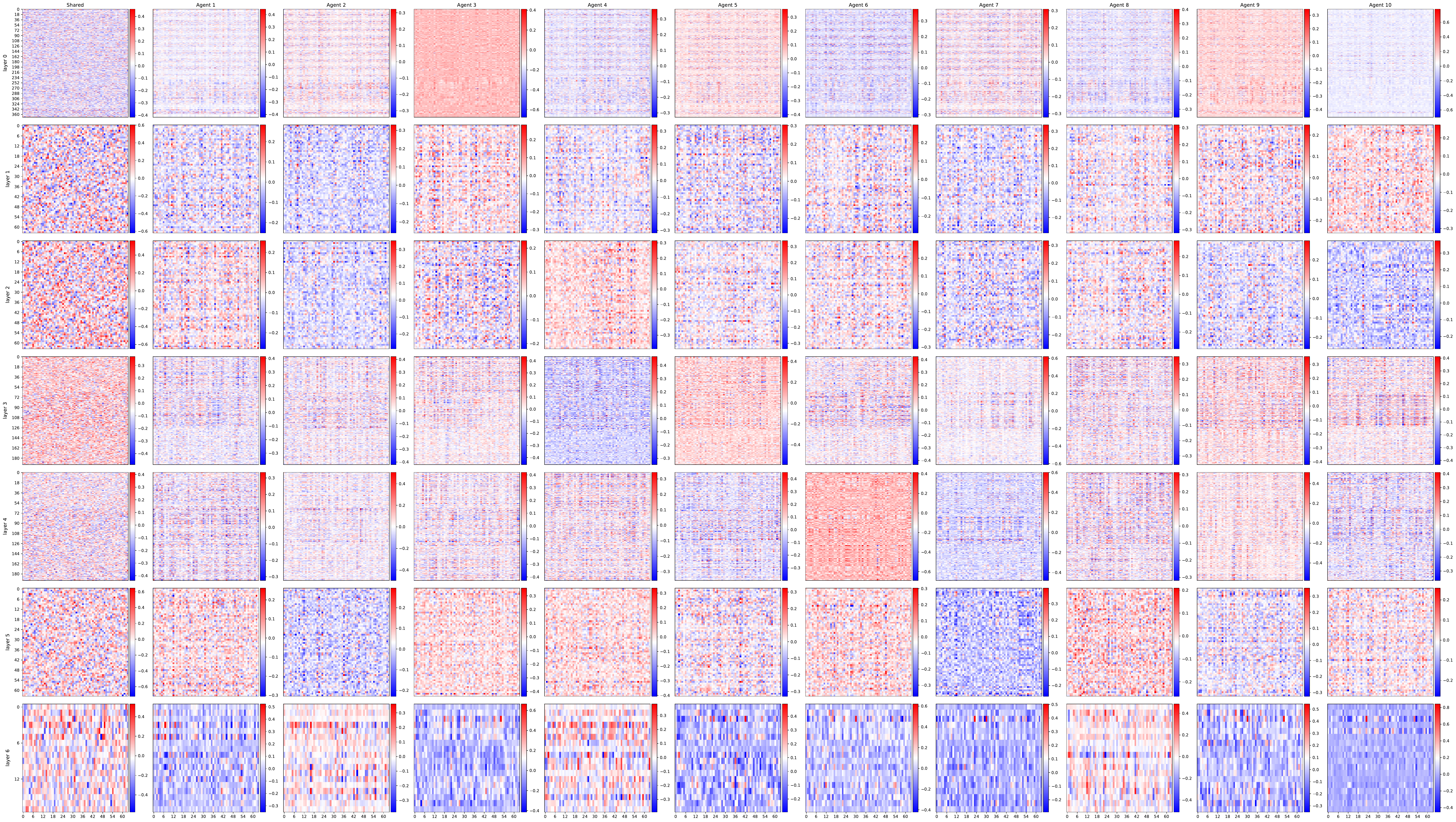}
\end{center}
\caption{Layer Activation of the map MMM2 using A2PO}
\label{fig:hm_a2po_mmm2}
\end{figure}

\begin{figure}
\begin{center}
\includegraphics[width=0.8\linewidth]{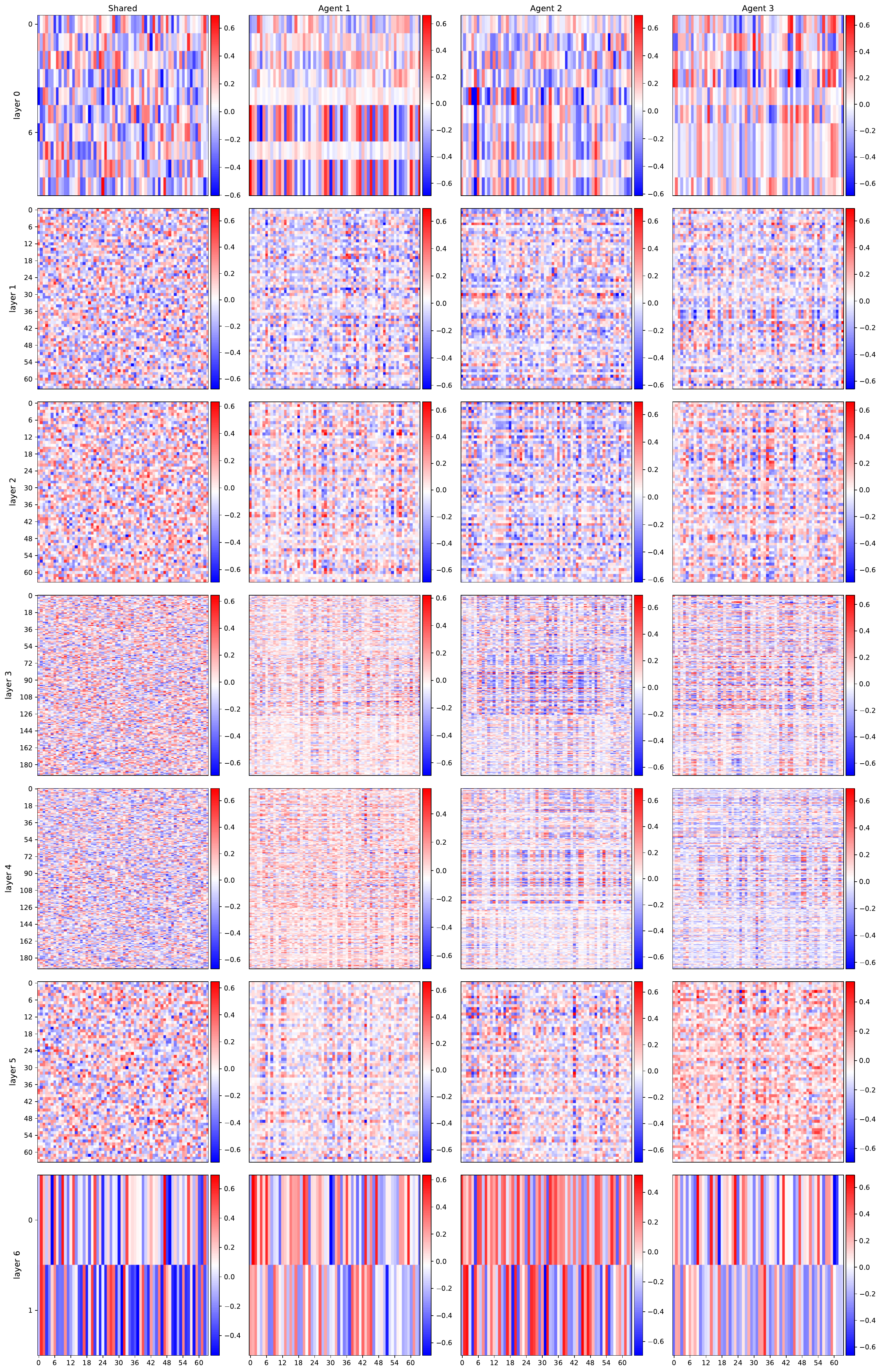}
\end{center}
\caption{Layer Activation of the map Walker3x2 using A2PO}
\label{fig:hm_a2po_walker3x2}
\end{figure}

\begin{figure}
\begin{center}
\includegraphics[width=\linewidth]{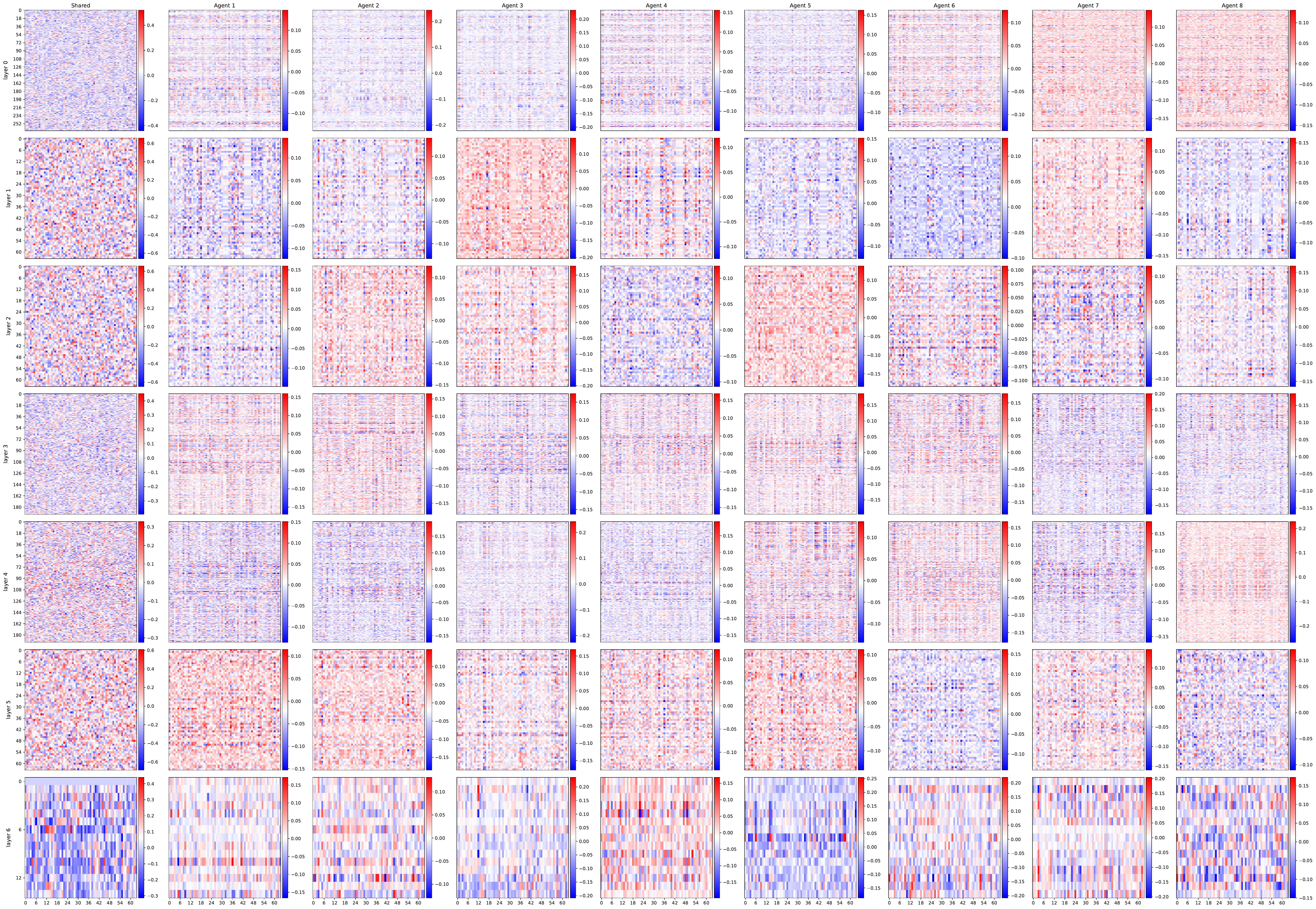}
\end{center}
\caption{Layer Activation of the map 3s5z\_vs\_3s6z using MAPPO}
\label{fig:hm_mappo_3s56z}
\end{figure}

\begin{figure}
\begin{center}
\includegraphics[width=\linewidth]{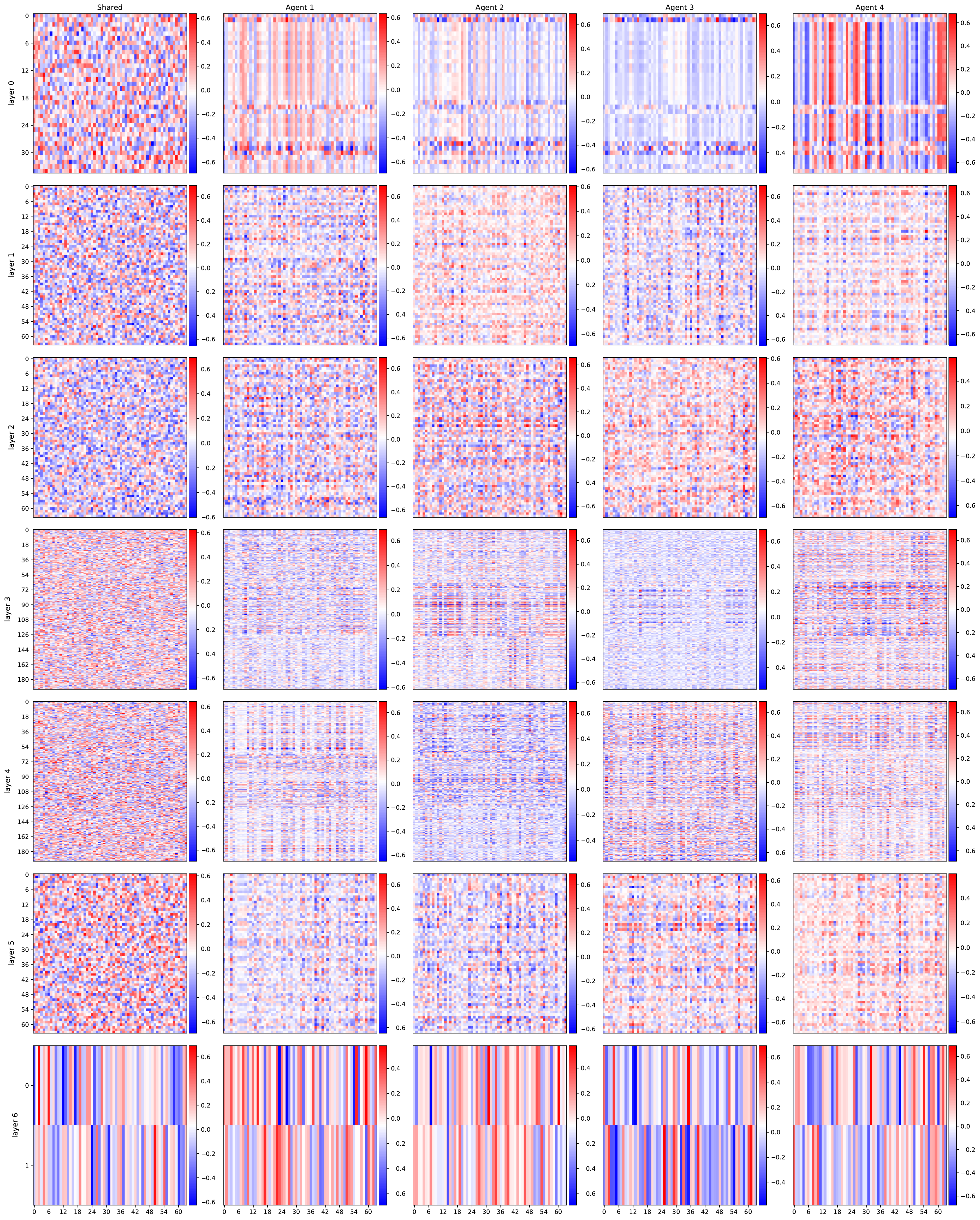}
\end{center}
\caption{Layer Activation of the map Ant4x2 using MAPPO}
\label{fig:hm_mappo_ant4x2}
\end{figure}

\begin{figure}
\begin{center}
\includegraphics[width=\linewidth]{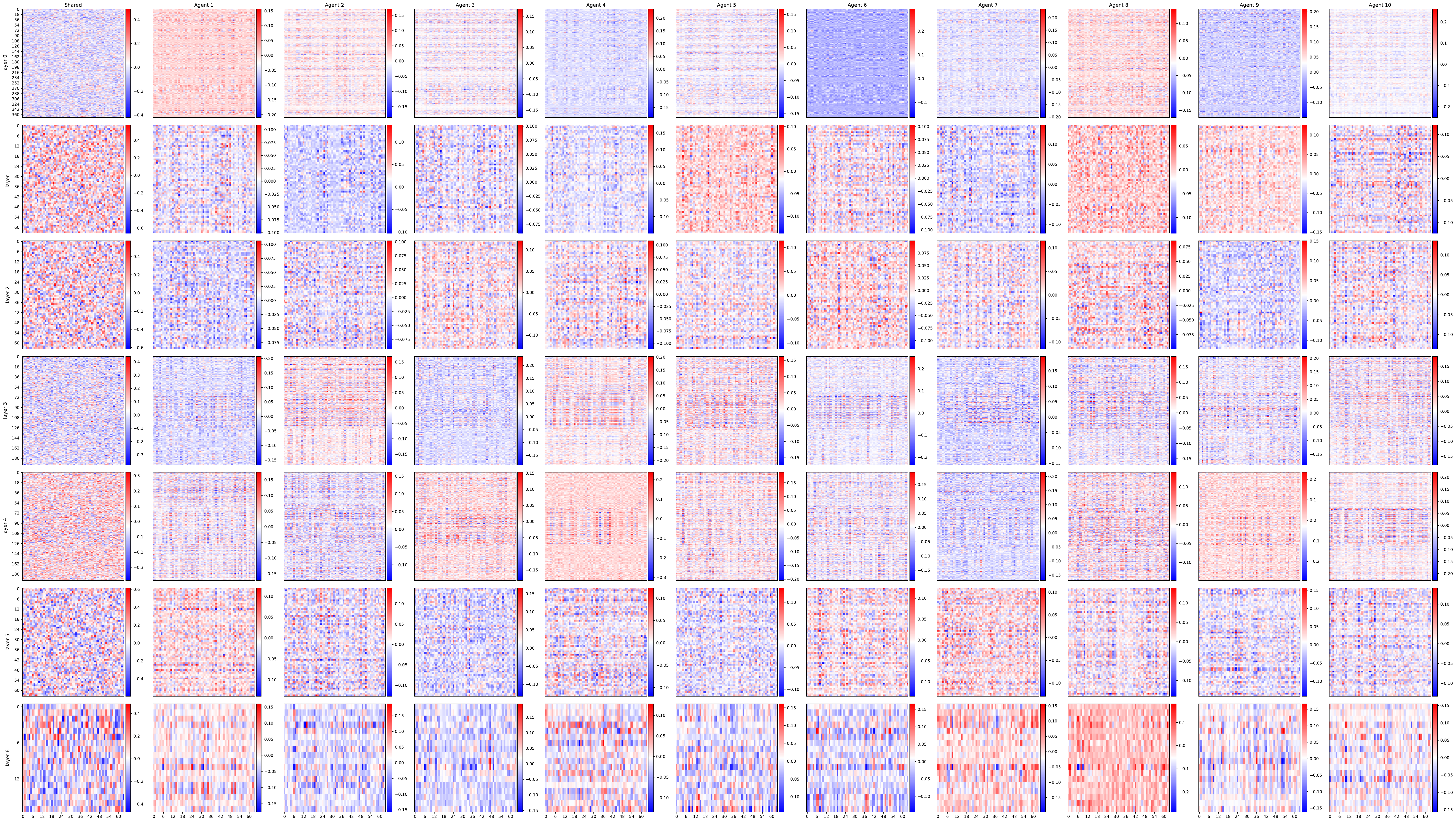}
\end{center}
\caption{Layer Activation of the map MMM2 using MAPPO}
\label{fig:hm_mappo_mmm2}
\end{figure}

\begin{figure}
\begin{center}
\includegraphics[width=0.85\linewidth]{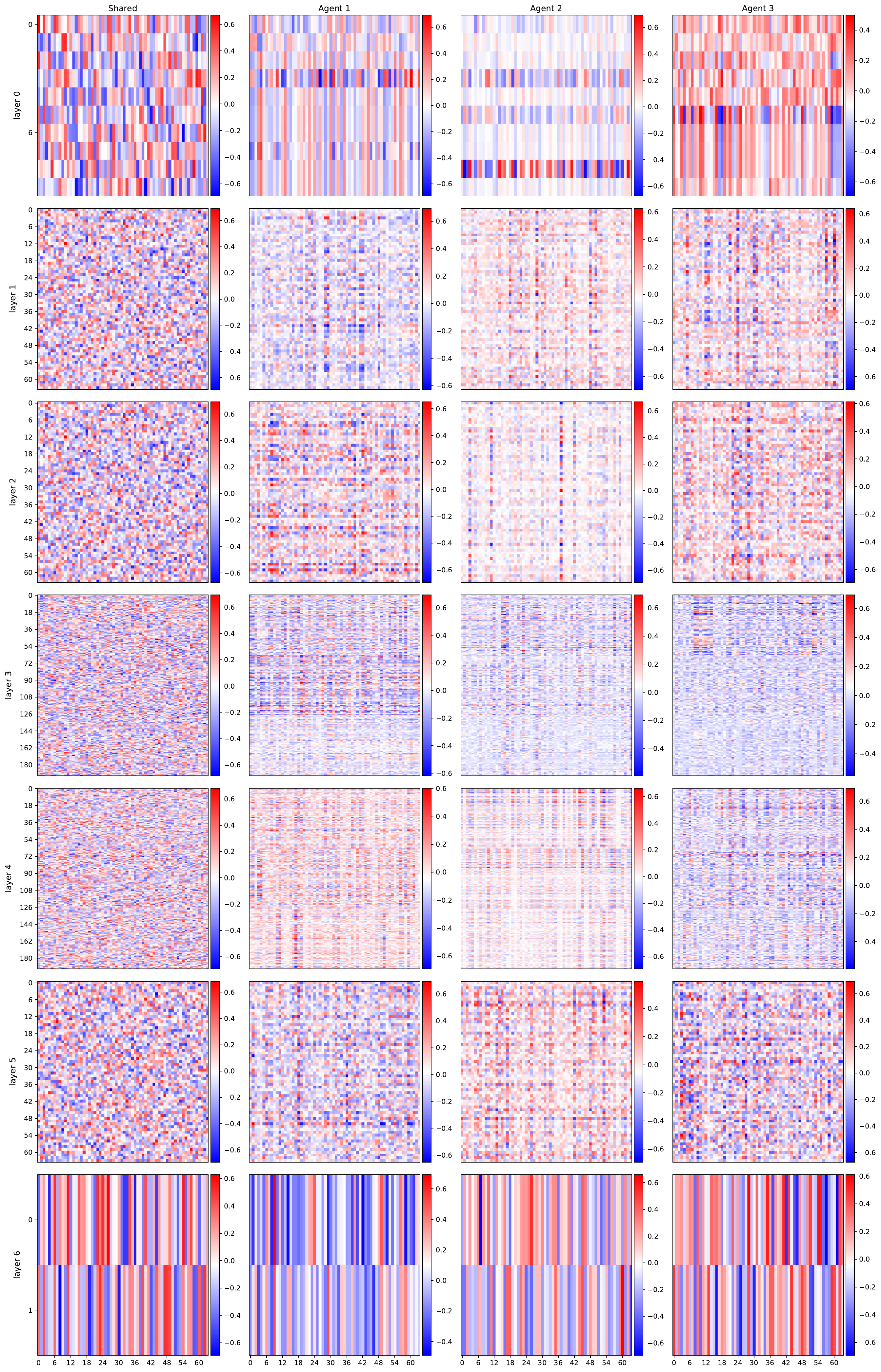}
\end{center}
\caption{Layer Activation of the map Walker3x2 using MAPPO}
\label{fig:hm_mappo_walker3x2}
\end{figure}

\begin{figure}
\begin{center}
\includegraphics[width=\linewidth]{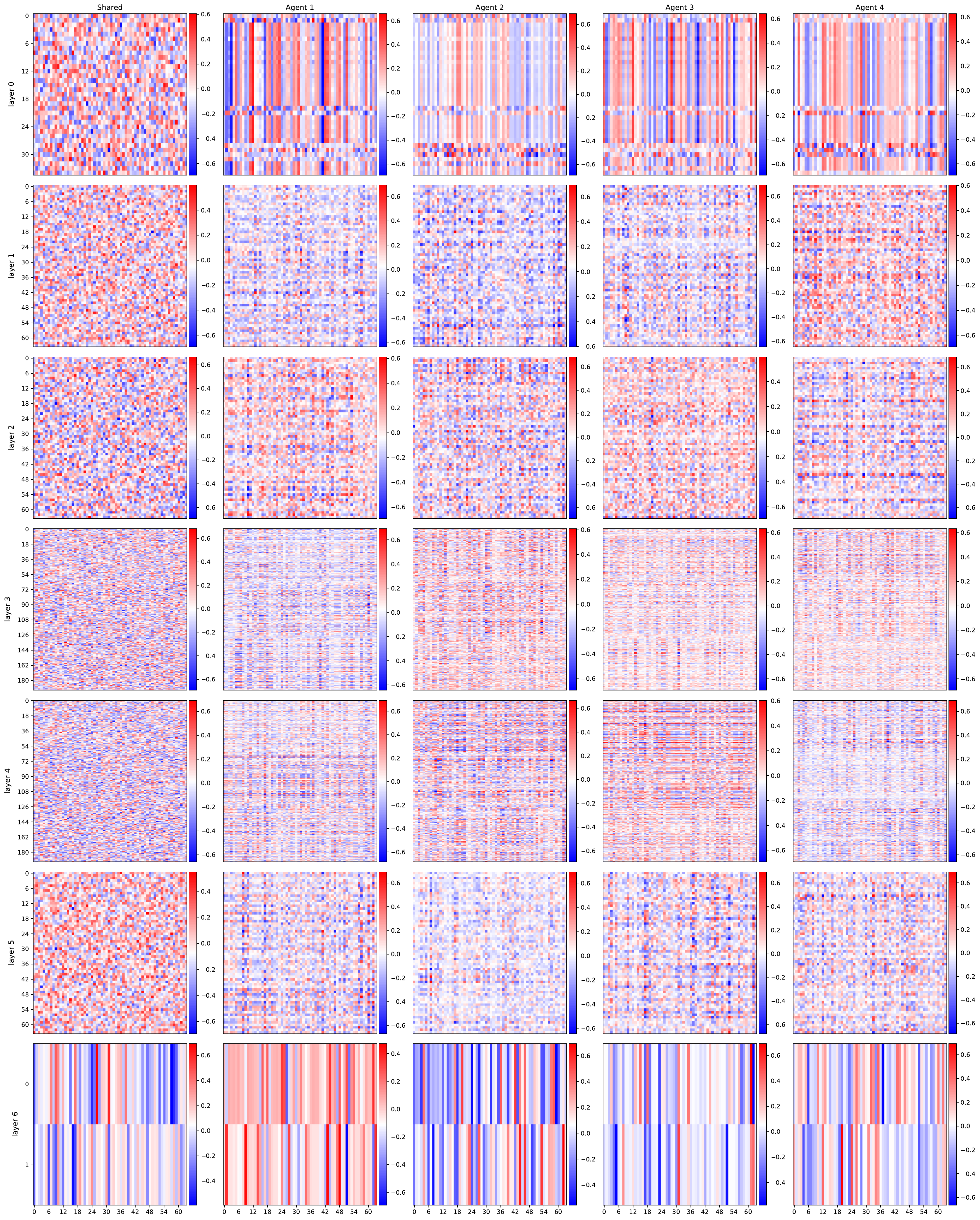}
\end{center}
\caption{Layer Activation of the map Ant4x2 using A2PO at 2M steps}
\label{fig:hm_a2po_ant4x2_2M}
\end{figure}

\begin{figure}
\begin{center}
\includegraphics[width=\linewidth]{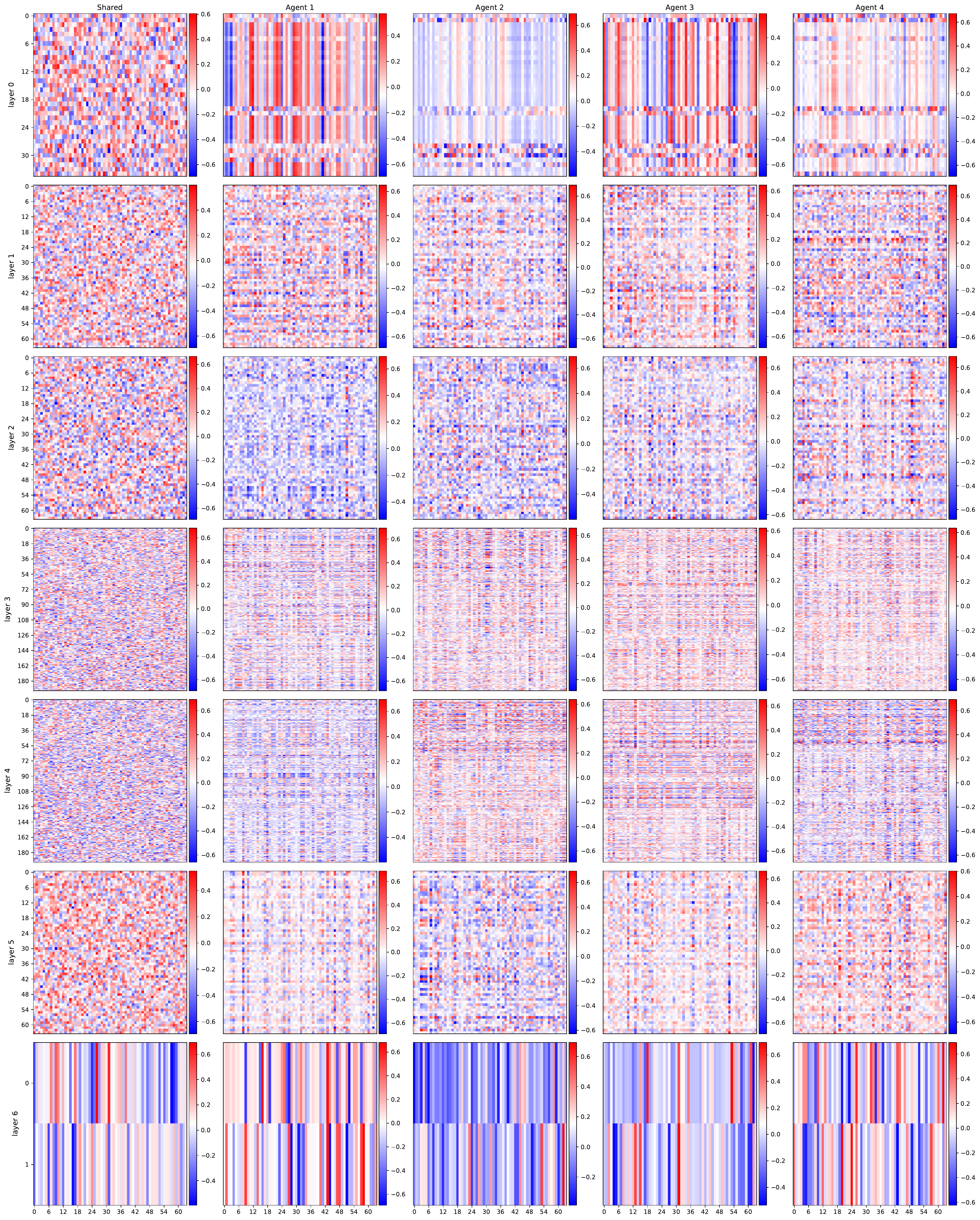}
\end{center}
\caption{Layer Activation of the map Ant4x2 using A2PO at 4M steps}
\label{fig:hm_a2po_ant4x2_4M}
\end{figure}

\begin{figure}
\begin{center}
\includegraphics[width=\linewidth]{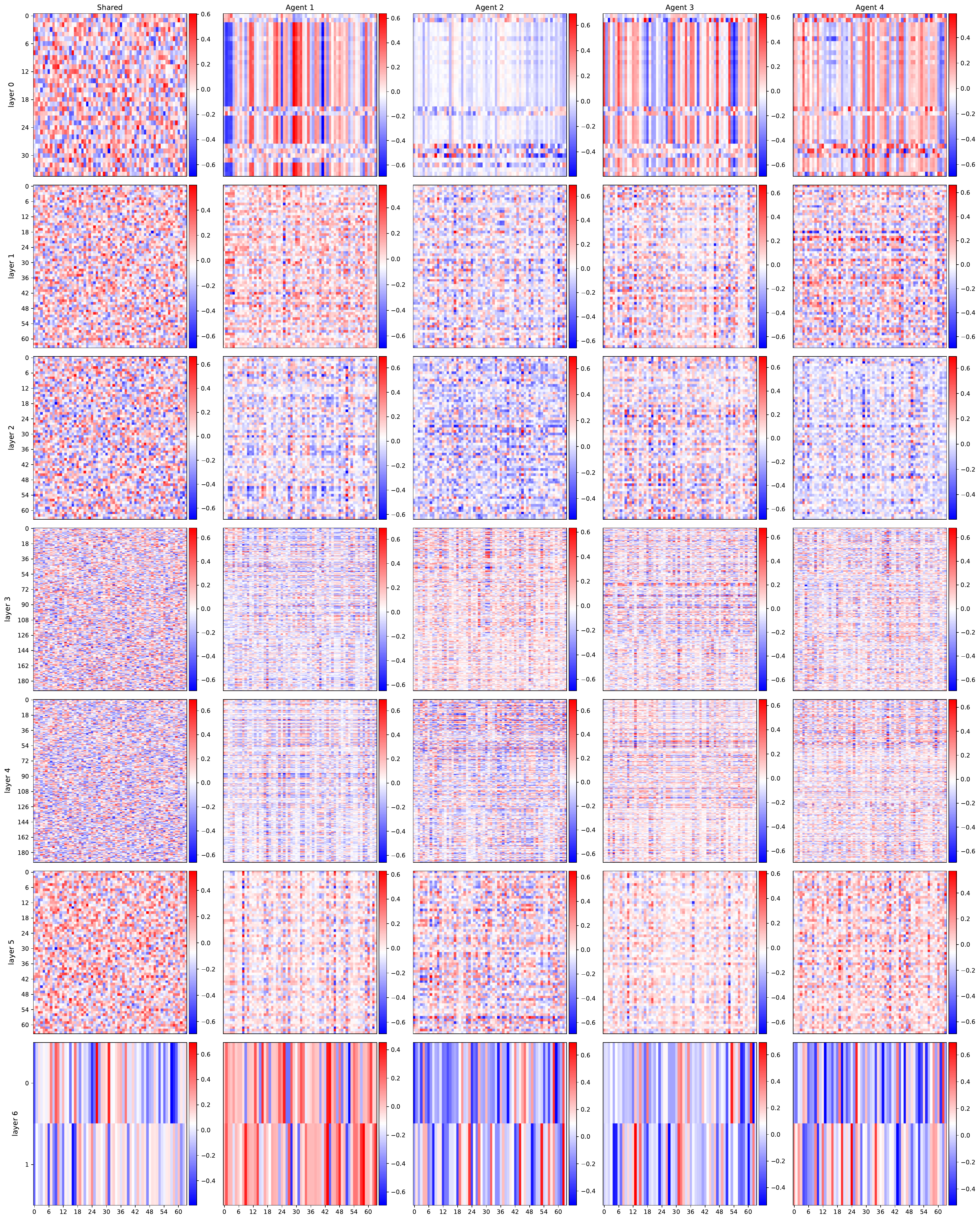}
\end{center}
\caption{Layer Activation of the map Ant4x2 using A2PO at 6M steps}
\label{fig:hm_a2po_ant4x2_6M}
\end{figure}

\begin{figure}
\begin{center}
\includegraphics[width=\linewidth]{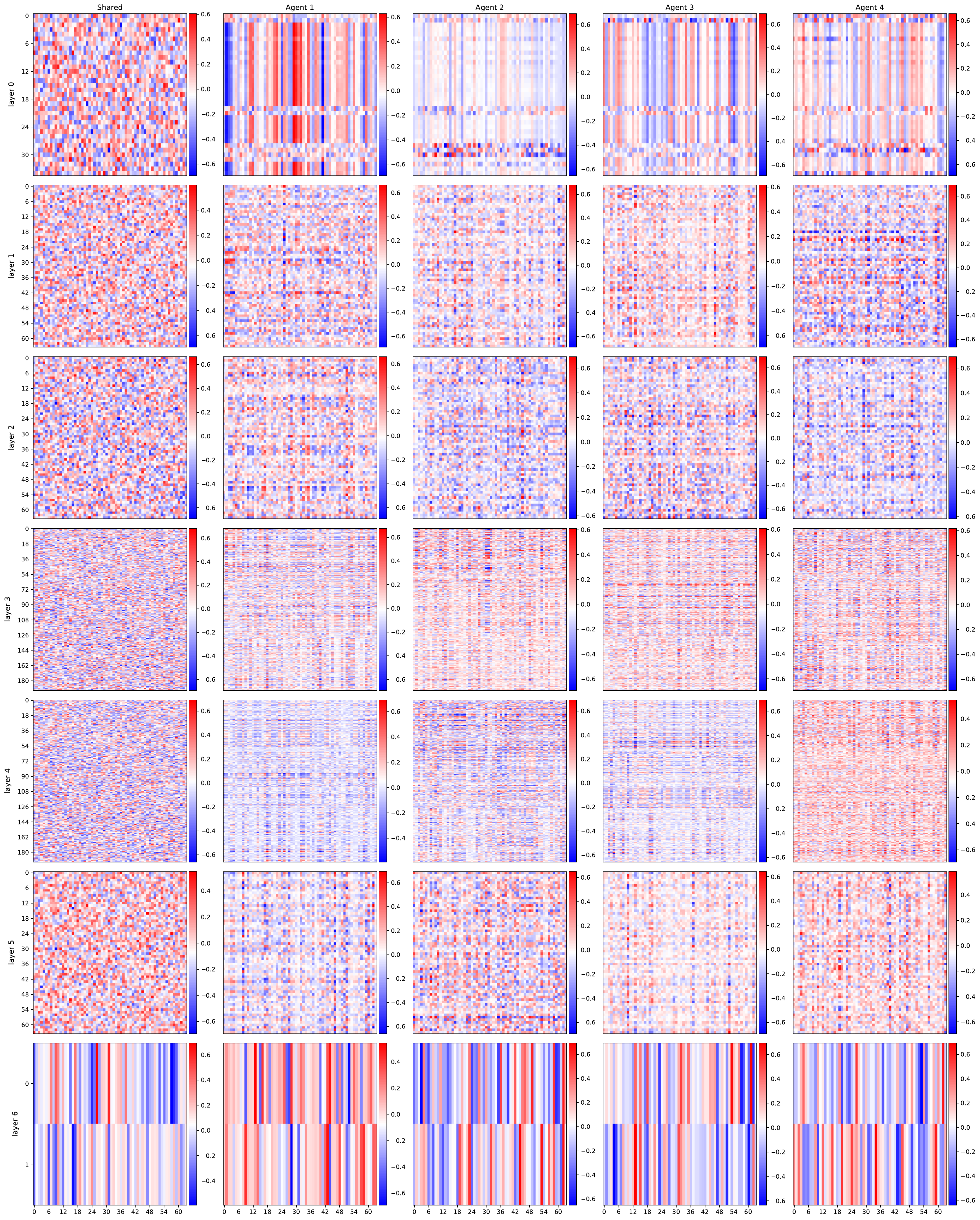}
\end{center}
\caption{Layer Activation of the map Ant4x2 using A2PO at 8M steps}
\label{fig:hm_a2po_ant4x2_8M}
\end{figure}

Figure~\ref{fig:sparsity_comparison} demonstrates the sparsity introduced by LoRASA in the policy parameter space. The percentage of policy parameters above various threshold values is significantly higher for the shared policy (\(\lvert\theta_\text{shared}\rvert\)) compared to LoRASA-adapted parameters (\(\lvert\delta\theta\rvert\)). This suggests that LoRASA fine-tuning effectively prunes the parameter space by focusing updates on a smaller, behaviorally critical subspace.

At lower thresholds, both shared and LoRA-adapted parameters maintain a higher proportion of active values. However, as the threshold increases, the LoRA curve drops off more sharply than the shared policy curve. This indicates that LoRA adaptations primarily influence low-magnitude adjustments, reinforcing its role as a lightweight mechanism for agent-specific fine-tuning without unnecessarily inflating parameter magnitudes.

This analysis ties the sparsity observation to LoRASA's broader benefits, reinforcing its practical and theoretical strengths in MARL.

\clearpage

\subsection{Hyperparameters}
\label{subsec:hyperparameters}

\begin{table}[h]
\centering
\begin{tabular}{lllcll}
          \toprule      & \multicolumn{2}{c}{A2PO PS+LoRA} && \multicolumn{2}{c}{A2PO SePS+LoRA} \\ \cmidrule{2-3} \cmidrule{5-6} 
Scenario        & r           & Checkpoint    &      & r            & Checkpoint          \\ \midrule
Halfcheetah 2x3 & 8           & 3.00E+06      &      & N/A          & N/A                 \\
Walker 3x2      & 8           & 2.00E+06       &     & 8            & 2.00E+06            \\
Ant 4x2         & 8           & 4.00E+06        &    & 8            & 4.00E+06            \\
Humanoid $9|8$    & 16          & 3.00E+06         &   & N/A          & N/A                 \\ \midrule
3s5z            & 8           & 2.00E+05          &  & 8            & 3.00E+05            \\
1c3s5z          & 8           & 5.00E+05           & & 8            & 1.00E+05            \\
3s5z\_vs\_3s6z  & 16          & 2.00E+06            && 16           & 5.00E+06            \\
MMM2            & 8           & 2.00E+06            && 16           & 2.00E+06 \\      \bottomrule    
\end{tabular}
\caption{Ranks and checkpoints for A2PO PS+LoRA and SePS+LoRA in MAMuJoCo and SMAC}
\label{tab:a2po_lora_hyperparams}
\end{table}

\begin{table}[h]
\centering
\begin{tabular}{lllcll}
     \toprule           & \multicolumn{2}{c}{MAPPO PS+LoRA} &  & \multicolumn{2}{c}{MAPPO SePS+LoRA} \\ \cmidrule{2-3} \cmidrule{5-6} 
Scenario        & r           & Checkpoint          &  & r             & Checkpoint          \\ \midrule
Halfcheetah 2x3 & 8           & 1.00E+05            &  & N/A           & N/A                 \\
Walker 3x2      & 8           & 2.00E+06            &  & 8             & 7.00E+06            \\
Ant 4x2         & 8           & 4.00E+06            &  & 8             & 4.00E+06            \\
Humanoid $9|8$    & 16          & 1.00E+06            &  & N/A           & N/A                 \\ \midrule
3s5z            & 8           & 2.00E+05            &  & 8             & 2.00E+05            \\
1c3s5z          & 8           & 5.00E+05            &  & 8             & 5.00E+05            \\
3s5z\_vs\_3s6z  & 8           & 4.00E+06            &  & 8             & 4.00E+06            \\
MMM2            & 8           & 7.00E+06            &  & 8             & 6.00E+06   \\ \bottomrule        
\end{tabular}
\caption{Ranks and checkpoints for MAPPO PS+LoRA and SePS+LoRA in MAMuJoCo and SMAC}
\label{tab:mappo_lora_hyperparams}
\end{table}

\begin{table}[h]
\centering
\begin{tabular}{ll}
\toprule
gamma                  & 0.99     \\
gain                   & 0.01     \\
activation             & ReLU     \\
optimizer              & Adam     \\
optim eps              & 1.00E-05 \\
MLP hidden layer       & 2        \\
hidden layer after RNN & 1        \\
actor network          & RNN      \\
chunk length           & 10       \\
max grad norm          & 10       \\
hidden layer dim       & 64       \\
num mini-batch         & 1       \\
\bottomrule
\end{tabular}
\caption{Global hyperparameters applicable across all environments, scenarios, algorithms, and parameter sharing methods.}
\label{tab:global_hyperparams}
\end{table}

\begin{table}[h]
\centering
\begin{tabular}{lllll}\toprule
                & Halfcheetah 2x3 & Walker 3x2 & Ant 4x2 & Humanoid $9|8$ \\ \midrule
episode length  & 4000            & 4000       & 4000    & 4000         \\
eval episode    & 10              & 10         & 10      & 10           \\
rollout threads & 16              & 16         & 25      & 25           \\
gae lambda      & 0.93            & 0.93       & 0.93    & 0.9         \\ \bottomrule
\end{tabular}
\caption{MAMuJoCo scenario-specific hyperparameters common to all algorithms and parameter sharing methods.}
\label{tab:mujoco_common_hyperparams}
\end{table}

\begin{table}[h]
\centering
\begin{tabular}{lllll}\toprule
                & 3s5z & 1c3s5z & 3s5z\_vs\_3s6z & MMM2 \\ \midrule
episode length  & 3200 & 3200   & 3200           & 3200 \\
eval episode    & 32   & 32     & 32             & 32   \\
rollout threads & 10   & 10     & 10             & 10   \\
gae lambda      & 0.95 & 0.95   & 0.9            & 0.95\\ \bottomrule
\end{tabular}
\caption{SMAC scenario-specific hyperparameters common to all algorithms and parameter sharing methods.}
\label{tab:smac_common_hyperparams}
\end{table}

\begin{table}[h]
\centering
\begin{tabular}{llllll}
\toprule Scenario        & clip & epoch & actor lr       & critic lr & entropy coefficient \\ \midrule
Halfcheetah 2x3 & 0.2  & 5     & 3.00E-04 & 3.00E-04  & 0                   \\
Walker 3x2      & 0.2  & 5     & 3.00E-04 & 3.00E-04  & 0                   \\
Ant 4x2         & 0.2  & 8     & 3.00E-04 & 3.00E-04  & 0                   \\
Humanoid $9|8$    & 0.2  & 5     & 3.00E-04 & 3.00E-04  & 0                   \\ \hline
3s5z            & 0.2  & 5     & 5.00E-04 & 5.00E-04  & 0.01                \\
1c3s5z          & 0.2  & 5     & 5.00E-04 & 5.00E-04  & 0.01                \\
3s5z\_vs\_3s6z  & 0.2  & 5     & 5.00E-04 & 5.00E-04  & 0.01                \\
MMM2            & 0.2  & 5     & 5.00E-04 & 5.00E-04  & 0.01       \\ \bottomrule        
\end{tabular}
\caption{Common hyperparameters for A2PO and MAPPO baselines, including NPS, MTL and SePS for A2PO, and PS, NPS, MTL, and SePS for MAPPO.}
\label{tab:algo_common_hyperparams}
\end{table}

\begin{table}[h]
\centering
\begin{tabular}{llllll}
\toprule Scenario        & clip & epoch & actor lr & critic lr & entropy coefficient \\ \midrule
Halfcheetah 2x3 & 0.2  & 5     & 3.00E-04 & 3.00E-04  & 0                   \\
Walker 3x2      & 0.2  & 3     & 3.00E-04 & 3.00E-04  & 0                   \\
Ant 4x2         & 0.2  & 3     & 3.00E-04 & 3.00E-04  & 0                   \\
Humanoid $9|8$    & 0.2  & 3     & 3.00E-04 & 3.00E-04  & 0                   \\ \midrule
3s5z            & 0.2  & 3     & 5.00E-04 & 5.00E-04  & 0.01                \\
1c3s5z          & 0.2  & 3     & 5.00E-04 & 5.00E-04  & 0.01                \\
3s5z\_vs\_3s6z  & 0.1  & 5     & 5.00E-04 & 5.00E-04  & 0.01                \\
MMM2            & 0.2  & 5     & 5.00E-04 & 5.00E-04  & 0.01     \\ \bottomrule          
\end{tabular}
\caption{Specific hyperparameters for A2PO PS.}
\label{tab:a2po_hyperparams_ps}
\end{table}

\begin{table}[h]
\centering
\begin{tabular}{llllll}
\toprule Scenario        & clip & epoch & actor lr & critic lr & entropy coefficient \\ \midrule
Halfcheetah 2x3 & 0.2  & 5     & 3.00E-04 & 3.00E-04  & 0                   \\
Walker 3x2      & 0.2  & 5     & 3.00E-04 & 3.00E-04  & 0                   \\
Ant 4x2         & 0.2  & 8     & 3.00E-04 & 3.00E-04  & 0                   \\
Humanoid $9|8$    & 0.2  & 5     & 3.00E-04 & 3.00E-04  & 0                   \\ \midrule
3s5z            & 0.2  & 5     & 5.00E-04 & 5.00E-04  & 0.01                \\
1c3s5z          & 0.2  & 3     & 5.00E-04 & 5.00E-04  & 0.01                \\
3s5z\_vs\_3s6z  & 0.2  & 5     & 5.00E-04 & 5.00E-04  & 0.01                \\
MMM2            & 0.2  & 5     & 5.00E-04 & 5.00E-04  & 0.01       \\ \bottomrule        
\end{tabular}
\caption{Specific hyperparameters for A2PO PS+LoRA and SePS+LoRA.}
\label{tab:a2po_hyperparams_pslora_sepslora}
\end{table}

\begin{table}[h]
\centering
\begin{tabular}{llllll}
\toprule Scenario        & clip & epoch & actor lr & critic lr & entropy coefficient \\ \midrule
Halfcheetah 2x3 & 0.2  & 5     & 3.00E-04 & 3.00E-04  & 0                   \\
Walker 3x2      & 0.2  & 5     & 3.00E-04 & 3.00E-04  & 0                   \\
Ant 4x2         & 0.2  & 8     & 3.00E-04 & 3.00E-04  & 0                   \\
Humanoid $9|8$    & 0.2  & 5     & 3.00E-04 & 3.00E-04  & 0                   \\ \midrule
3s5z            & 0.2  & 5     & 3.00E-04 & 3.00E-04  & 0.01                \\
1c3s5z          & 0.2  & 5     & 3.00E-04 & 3.00E-04  & 0.01                \\
3s5z\_vs\_3s6z  & 0.05 & 5     & 3.00E-04 & 3.00E-04  & 0.001               \\
MMM2            & 0.05 & 5     & 3.00E-04 & 3.00E-04  & 0.001    \\ \bottomrule          
\end{tabular}
\caption{Specific hyperparameters for MAPPO PS+LoRA and SePS+LoRA.}
\label{tab:mappo_hyperparams_pslora_sepslora}
\end{table}

\begin{table}[h]
\centering
\resizebox{\textwidth}{!}{%
\begin{tabular}{llclllclllclllclll}
\toprule
\multicolumn{1}{c}{} & \multicolumn{1}{c}{PS} &  & \multicolumn{3}{c}{$2\times10^6$} &  & \multicolumn{3}{c}{$4\times10^6$} &  & \multicolumn{3}{c}{$6\times10^6$} &  & \multicolumn{3}{c}{$8\times10^6$} \\ \cmidrule{2-2} \cmidrule{4-6} \cmidrule{8-10} \cmidrule{12-14} \cmidrule{16-18} 
Layer &
  \multicolumn{1}{c}{$\|\theta\|_F$} &
   &
  \multicolumn{1}{c}{$\|\theta_{shared}\|_F$} &
  \multicolumn{1}{c}{$\|\theta'\|_F$} &
  \multicolumn{1}{c}{$\|\delta \theta\|_F$} &
   &
  \multicolumn{1}{c}{$\|\theta_{shared}\|_F$} &
  \multicolumn{1}{c}{$\|\theta'\|_F$} &
  \multicolumn{1}{c}{$\|\delta \theta\|_F$} &
   &
  \multicolumn{1}{c}{$\|\theta_{shared}\|_F$} &
  \multicolumn{1}{c}{$\|\theta'\|_F$} &
  \multicolumn{1}{c}{$\|\delta \theta\|_F$} &
   &
  \multicolumn{1}{c}{$\|\theta_{shared}\|_F$} &
  \multicolumn{1}{c}{$\|\theta'\|_F$} &
  \multicolumn{1}{c}{$\|\delta \theta\|_F$} \\ \midrule
Linear 1             & 8.97                   &  & 8.55   & 8.78   & 1.87 &  & 8.70   & 8.82   & 1.48 &  & 8.85   & 8.89   & 1.04 &  & 8.89   & 8.91   & 0.59 \\
Linear 2             & 12.13                  &  & 11.55  & 11.84  & 2.64 &  & 11.80  & 11.94  & 1.90 &  & 11.97  & 12.03  & 1.30 &  & 12.06  & 12.10  & 0.90 \\
Linear 3             & 12.08                  &  & 11.57  & 11.84  & 2.46 &  & 11.82  & 11.99  & 1.96 &  & 11.93  & 12.01  & 1.30 &  & 12.01  & 12.04  & 0.91 \\
GRU x proj 4         & 11.14                  &  & 9.08   & 10.64  & 5.47 &  & 10.10  & 10.81  & 3.78 &  & 10.43  & 10.78  & 2.67 &  & 10.79  & 10.93  & 1.75 \\
GRU h proj 5         & 10.99                  &  & 8.96   & 10.36  & 5.10 &  & 9.81   & 10.59  & 3.92 &  & 10.22  & 10.54  & 2.52 &  & 10.56  & 10.75  & 1.85 \\
Linear 6             & 11.99                  &  & 11.57  & 11.74  & 2.05 &  & 11.80  & 11.88  & 1.56 &  & 11.88  & 11.92  & 0.98 &  & 11.93  & 11.95  & 0.70 \\
Linear 7             & 0.25                   &  & 0.16   & 0.33   & 0.30 &  & 0.21   & 0.32   & 0.27 &  & 0.22   & 0.29   & 0.18 &  & 0.25   & 0.26   & 0.11 \\ \bottomrule
\end{tabular}%
}
\caption{Norms of weights in different layers of the neural network for MAPPO PS and PS+LoRA with different checkpoints in Ant 4x2: including the norms of weights of PS at $12 \times 10^6$ steps ($\theta$), the norms of weights of pretrained PS ($\theta_{shared}$), the average norms of merged weights across agents ($\theta'$), and the average norms of weights for LoRA adapters ($\delta \theta$).}
\label{tab:norms_ckpt_mappo_ant4x2}
\end{table}

\begin{table}[h]
\centering
\resizebox{\textwidth}{!}{%
\begin{tabular}{llclcllllcllll}
\toprule
             &       &  &       &  & \multicolumn{4}{c}{$\|\theta'\|_F$} &  & \multicolumn{4}{c}{$\|\delta \theta\|_F$} \\ \cmidrule{6-9} \cmidrule{11-14} 
Layer &
  \multicolumn{1}{c}{$\|\theta \|_F$} &
   &
  \multicolumn{1}{c}{$\|\theta_{shared}\|_F$} &
   &
  \multicolumn{1}{c}{$r=4$} &
  \multicolumn{1}{c}{$r=8$} &
  \multicolumn{1}{c}{$r=16$} &
  \multicolumn{1}{c}{$r=64$} &
   &
  \multicolumn{1}{c}{$r=4$} &
  \multicolumn{1}{c}{$r=8$} &
  \multicolumn{1}{c}{$r=16$} &
  \multicolumn{1}{c}{$r=64$} \\ \midrule
Linear 1     & 8.97  &  & 8.74  &  & 8.86   & 8.82  & 8.85  & 8.98  &  & 1.37    & 1.48    & 1.44    & 1.98   \\
Linear 2     & 12.13 &  & 11.86 &  & 12.00  & 11.94 & 12.02 & 12.28 &  & 1.73    & 1.90    & 1.98    & 2.86   \\
Linear 3     & 12.08 &  & 11.86 &  & 12.03  & 11.99 & 12.05 & 12.27 &  & 1.96    & 1.96    & 2.07    & 2.91   \\
GRU x proj 4 & 11.14 &  & 10.03 &  & 10.76  & 10.81 & 10.97 & 13.01 &  & 3.85    & 3.78    & 4.42    & 7.99   \\
GRU h proj 5 & 10.99 &  & 9.81  &  & 10.61  & 10.59 & 10.85 & 13.79 &  & 3.95    & 3.92    & 4.38    & 9.39   \\
Linear 6     & 11.99 &  & 11.78 &  & 11.94  & 11.88 & 11.88 & 12.13 &  & 1.74    & 1.56    & 1.57    & 2.68   \\
Linear 7     & 0.25  &  & 0.21  &  & 0.31   & 0.32  & 0.30  & 0.25  &  & 0.25    & 0.27    & 0.22    & 0.19   \\ \bottomrule
\end{tabular}%
}
\caption{Norms of weights in different layers of the neural network for MAPPO PS and different ranks of PS+LoRA fine tuning in Ant 4x2: including the norms of weights of PS at $12 \times 10^6$ steps ($\theta$), the norms of weights of pretrained PS at $4 \times 10^6$ steps ($\theta_{shared}$), the average norms of merged weights across agents ($\theta'$), and the average norms of weights for LoRA adapters ($\delta \theta$).}
\label{tab:norms_r_mappo_ant4x2}
\end{table}

\begin{table}[h]
\centering
\resizebox{\textwidth}{!}{%
\begin{tabular}{llclllclllclllclll}\toprule
             & \multicolumn{1}{c}{} &  & \multicolumn{3}{c}{$1\times10^6$} &  & \multicolumn{3}{c}{$3\times10^6$} &  & \multicolumn{3}{c}{$5\times10^6$} &  & \multicolumn{3}{c}{$7\times10^6$} \\ \cmidrule{4-6} \cmidrule{8-10} \cmidrule{12-14} \cmidrule{16-18} 
Layer &
  \multicolumn{1}{c}{$\|\theta \|_F$} &
   &
  \multicolumn{1}{c}{$\|\theta_{shared}\|_F$} &
  \multicolumn{1}{c}{$\|\theta'\|_F$} &
  \multicolumn{1}{c}{$\|\delta \theta\|_F$} &
   &
  \multicolumn{1}{c}{$\|\theta_{shared}\|_F$} &
  \multicolumn{1}{c}{$\|\theta'\|_F$} &
  \multicolumn{1}{c}{$\|\delta \theta\|_F$} &
   &
  \multicolumn{1}{c}{$\|\theta_{shared}\|_F$} &
  \multicolumn{1}{c}{$\|\theta'\|_F$} &
  \multicolumn{1}{c}{$\|\delta \theta\|_F$} &
   &
  \multicolumn{1}{c}{$\|\theta_{shared}\|_F$} &
  \multicolumn{1}{c}{$\|\theta'\|_F$} &
  \multicolumn{1}{c}{$\|\delta \theta\|_F$} \\ \midrule
Linear 1     & 17.32                &  & 11.70  & 12.76  & 5.15 &  & 12.95  & 13.79  & 4.70 &  & 13.81  & 14.48  & 4.33 &  & 15.11  & 15.49  & 3.39 \\
Linear 2     & 12.69                &  & 11.38  & 11.57  & 2.02 &  & 11.64  & 11.80  & 1.93 &  & 11.85  & 11.99  & 1.83 &  & 12.18  & 12.26  & 1.37 \\
Linear 3     & 12.82                &  & 11.38  & 11.59  & 2.20 &  & 11.64  & 11.79  & 1.95 &  & 11.85  & 12.01  & 1.98 &  & 12.22  & 12.30  & 1.46 \\
GRU x proj 4 & 13.93                &  & 8.41   & 9.69   & 4.80 &  & 9.61   & 10.56  & 4.30 &  & 10.52  & 11.31  & 4.13 &  & 11.86  & 12.25  & 2.96 \\
GRU h proj 5 & 13.96                &  & 8.45   & 9.83   & 4.96 &  & 9.56   & 10.47  & 4.20 &  & 10.29  & 11.22  & 4.32 &  & 11.73  & 12.17  & 3.15 \\
Linear 6     & 13.02                &  & 11.40  & 11.60  & 2.18 &  & 11.67  & 11.83  & 1.97 &  & 11.92  & 12.10  & 2.02 &  & 12.29  & 12.39  & 1.64 \\
Linear 7     & 6.25                 &  & 4.61   & 6.09   & 2.96 &  & 5.25   & 5.70   & 1.52 &  & 5.50   & 5.87   & 1.34 &  & 5.82   & 5.97   & 0.93 \\ \bottomrule
\end{tabular}%
}
\caption{Norms of weights in different layers of the neural network for MAPPO PS and PS+LoRA with different checkpoints in MMM2: including the norms of weights of PS at $12\times10^6$ steps ($\theta$), the norms of weights of pretrained PS ($\theta_{shared}$), the average norms of merged weights across agents ($\theta'$), and the average norms of weights for LoRA adapters ($\delta \theta$).}
\label{tab:norms_ckpt_mappo_mmm2}
\end{table}

\begin{table}[h]
\centering
\resizebox{\textwidth}{!}{%
\begin{tabular}{llclcllllcllll}\toprule
             &       &  &       &  & \multicolumn{4}{c}{$\|\theta'\|_F$} &  & \multicolumn{4}{c}{$\|\delta \theta\|_F$} \\ \cmidrule{6-9} \cmidrule{11-14} 
Layer &
  \multicolumn{1}{c}{$\|\theta \|_F$} &
   &
  \multicolumn{1}{c}{$\|\theta_{shared}\|_F$} &
   &
  \multicolumn{1}{c}{$r=4$} &
  \multicolumn{1}{c}{$r=8$} &
  \multicolumn{1}{c}{$r=16$} &
  \multicolumn{1}{c}{$r=64$} &
   &
  \multicolumn{1}{c}{$r=4$} &
  \multicolumn{1}{c}{$r=8$} &
  \multicolumn{1}{c}{$r=16$} &
  \multicolumn{1}{c}{$r=64$} \\ \midrule
Linear 1     & 17.32 &  & 15.11 &  & 15.48  & 15.49 & 15.49 & 15.85 &  & 3.33    & 3.39    & 3.46    & 4.64   \\
Linear 2     & 12.69 &  & 12.18 &  & 12.26  & 12.26 & 12.25 & 12.36 &  & 1.40    & 1.37    & 1.39    & 1.87   \\
Linear 3     & 12.82 &  & 12.22 &  & 12.32  & 12.30 & 12.31 & 12.41 &  & 1.50    & 1.46    & 1.50    & 1.94   \\
GRU x proj 4 & 13.93 &  & 11.86 &  & 12.27  & 12.25 & 12.29 & 12.71 &  & 3.03    & 2.96    & 3.12    & 4.33   \\
GRU h proj 5 & 13.96 &  & 11.73 &  & 12.16  & 12.17 & 12.21 & 12.72 &  & 3.09    & 3.15    & 3.23    & 4.58   \\
Linear 6     & 13.02 &  & 12.29 &  & 12.38  & 12.39 & 12.40 & 12.50 &  & 1.58    & 1.64    & 1.67    & 2.14   \\
Linear 7     & 6.25  &  & 5.82  &  & 5.96   & 5.97  & 6.01  & 6.00  &  & 0.91    & 0.93    & 1.07    & 1.11   \\ \bottomrule
\end{tabular}%
}
\caption{Norms of weights in different layers of the neural network for MAPPO PS and different ranks of PS+LoRA fine tuning in MMM2: including the norms of weights of PS at $12 \times 10^6$ steps ($\theta$), the norms of weights of pretrained PS at $7 \times 10^6$ steps ($\theta_{shared}$), the average norms of merged weights across agents ($\theta'$), and the average norms of weights for LoRA adapters ($\delta \theta$).}
\label{tab:norms_r_mappo_mmm2}
\end{table}


\end{document}